\documentclass[aps,showpacs,showkeys,11pt]{revtex4}
\usepackage{epsfig}
\usepackage{amsmath, amssymb}
\newcommand{\beq}{\begin{equation}}
\newcommand{\eeq}{\end{equation}}
\newcommand{\beqa}{\begin{eqnarray}}
\newcommand{\eeqa}{\end{eqnarray}}
\def\GeV{\nobreak\,\mbox{GeV}}

\begin{document}
\title{Production and evolution path of dileptons at HADES energies}

\author{K. Schmidt$^{1}$, E. Santini$^{1}$, S. Vogel$^{1}$, C. Sturm$^{2}$, M. Bleicher$^{1}$ and H. St\"ocker$^{1,3,4}$}

\affiliation{\vspace*{.5cm}
$^1$Institut f\"ur Theoretische Physik, Goethe-Universit\"at,\\ 
Max-von-Laue-Str.~1, D-60438 Frankfurt am Main, Germany\\
$^2$Institut f\"ur Kernphysik, Goethe-Universit\"at,\\ 
Max-von-Laue-Str.~1, D-60438 Frankfurt am Main, Germany\\
$^3$Frankfurt Institute for Advanced Studies (FIAS),\\ 
Ruth-Moufang-Str.~1, D-60438 Frankfurt am Main, Germany\\
$^4$GSI - Helmholtzzentrum f\"ur Schwerionenforschung GmbH,\\
Planckstrasse~1, D-64291 Darmstadt, Germany
}

\date{\today}

\begin{abstract}
Dilepton production in intermediate energy nucleus-nucleus collisions as well as in 
elementary proton-proton reactions is analysed within the UrQMD transport model.
For C+C collisions at 1 AGeV and 2 AGeV the resulting 
invariant mass spectra are compared to recent HADES data. 
We find that the experimental spectrum for C+C at 2 AGeV is 
slightly overestimated by the theoretical calculations in the 
region around the vector meson peak, but fairly described 
in the low mass region, where the data is satisfactorily saturated 
by the Dalitz decay of the 
$\eta$ meson and the 
$\Delta$ resonance. 
At 1 AGeV an underestimation of the experimental 
data is found, pointing that at lower energies the low mass 
region is not fully saturated by standardly parametrized $\Delta$ Dalitz 
decays alone.
Furthermore, predictions for dilepton spectra for $pp$ reactions at 
1.25 GeV, 2.2 GeV and 3.5 GeV and Ar+KCl reactions at 1.75 AGeV
are presented. 
The study is complemented by a detailed investigation 
of the role of absorption of the parent particles 
on the corresponding dilepton yields in the regime which has so far been 
probed by HADES.
\end{abstract}
\pacs{13.40.Hq, 24.10.Lx, 25.75.-q, 25.75.Dw}
\keywords{Electromagnetic decays, 	Monte Carlo simulations, Relativistic heavy-ion collisions, Particle and resonance production} 
\maketitle

\section{Introduction}
In the last decades large experimental and theoretical efforts have 
been directed 
to the investigation of dilepton production in heavy ion collisions 
\cite{Mazzoni:1994rb,Li:1994cj,Agakishiev:1995xb,Li:1995qm,Rapp:1995zy,
Ko:1996is,Li:1996mi,Rapp:1997fs,Friman:1997tc,
Cassing:1997jz,Porter:1997rc,Bratkovskaya:1997mp,Ernst:1997yy,Shekhter:2003xd,
Arnaldi:2006jq,Adamova:2006nu,Agakichiev:2006tg,Agakishiev:2007ts,Cozma:2006vp,Schumacher:2006wc,
Thomere:2007cj,Ruppert:2007cr,Bratkovskaya:2007jk,Vogel:2007yu,Santini:2008pk}. 
Dileptons represent a particularly clean and penetrating probe of the 
hot and dense nuclear matter due to the fact that, once produced, they 
essentially do not interact with the surrounding hadronic matter. 
The analysis of the electromagnetic response of the dense and hot medium 
is tightly connected to the investigation of the in-medium modification 
of the vector meson properties. Vector mesons can directly decay into a 
lepton-antilepton pair. One therefore aims to infer information on the 
modifications induced by the medium on specific properties of the vector 
meson, such as its mass and/or its width, from the invariant mass dilepton 
spectra. 

A first generation of ultra-relativistic heavy ion collision experiments 
performed in the 
nineties observed an enhancement of dilepton production in heavy system at low invariant mass as 
compared to conventional hadronic cocktails and models 
\cite{Agakishiev:1995xb,Mazzoni:1994rb}. 
The enhancement could 
be later explained by the inclusion of an in-medium modified $\rho$ meson. 
At that time two possible scenarios, a dropping of the $\rho$ meson mass 
according to the Brown-Rho scaling hypothesis \cite{Brown:1991kk} 
and the Hatsuda and Lee sum rule prediction  
\cite{Hatsuda:1991ez}, or a ``melting'' of its 
spectral function as expected within many-body hadronic models 
\cite{Rapp:1997fs,Friman:1997tc,Peters:1997va,Lutz:2001mi}, 
have been offered in attempt to 
explain these data \cite{Li:1995qm,Ko:1996is,Li:1996mi,Rapp:1995zy,Cassing:1997jz}.  
If on the one side these experiments clearly showed the need for an inclusion 
of in-medium effects, on the 
other side it could not be decided, on the basis of the experimental data, 
whether the additional strength at lower invariant masses was due to a 
dropping of the vector meson mass or to the broadening of its spectral 
function. A first answer in this direction came from the measurements 
performed by the NA60 Collaboration \cite{Arnaldi:2006jq}. 
The data strongly favour the 
broadening over the dropping mass scenario. A similar 
conclusion is suggested by recent higher resolution CERES data 
\cite{Adamova:2006nu}.

At lower bombarding energies dileptons have been measured 
by the DLS Collaboration at BEVELAC \cite{Porter:1997rc}. 
The most striking result of the DLS experiment was an observed 
enhancement at lower invariant masses in nucleus-nucleus collisions at 1 AGeV 
with respect to the corresponding theoretical spectra resulting from 
transport calculations 
\cite{Bratkovskaya:1997mp,Ernst:1997yy,Shekhter:2003xd}. 
Differently to the ultra-relativistic 
case, none of the in-medium scenarios which had successfully explained the 
ultra-relativistic heavy ion collision data could account for the observed enhancement 
\cite{Bratkovskaya:1997mp,Ernst:1997yy} (this is known as the DLS puzzle). 
In the meanwhile the HADES spectrometer has been built at GSI 
with the aim of performing a systematic study of dilepton production 
in elementary, as well as heavy ion reactions. 
First HADES data have recently 
been presented \cite{Agakichiev:2006tg,Agakishiev:2007ts}, 
accompanied by a growing related theoretical activity 
\cite{Cozma:2006vp,Schumacher:2006wc,Thomere:2007cj,
Bratkovskaya:2007jk,Santini:2008pk}.

Aim of this work is a detailed investigation of dilepton production in 
heavy ion and elementary reactions at SIS energies. The analysis 
is performed within the microscopic UrQMD model,  a non-equilibrium transport approach
based on the quantum molecular dynamics
concept
\cite{aichelin86a,bass95c,winckelmann96a}.
The model allows for the production of all
established meson and baryon resonances up to
about $2\GeV$ with all
corresponding isospin projections  and antiparticle-states.
The collision term
describes particle production by 
resonant excitation channels
and, for higher energies, within a string fragmentation
scheme. For dilepton production at SIS 
energies, the resonant production of neutral
mesons is most important.
 The
model allows to study the full space time evolution for all hadrons, 
resonances and
their decay products. This permits to explore the emission patterns of 
the resonances
in detail and to gain insight into the origin of the resonances. 
UrQMD has been
successfully applied to study light and heavy ion reactions at SIS. 
Detailed
comparisons of UrQMD with a large body of experimental data at 
SIS energies can be
found in \cite{Sturm:2000dm}. 
For further details of the model the reader is referred
to \cite{Bass:1998ca,Bleicher:1999xi}, the latest version (v2.3) is 
described in \cite{Petersen:2008kb}.

The systems analysed here have been 
chosen according to 
the HADES program.
For those systems for which the HADES data and detector filter function 
are available a direct comparison to the data is performed. 
The additional calculations are given as predictions 
which can be compared to experimental data in the near future. 
The outline of the paper is the following: After a brief survey of the 
UrQMD model and 
of the therein implemented dilepton
production channels in Section \ref{The model}, 
proton-proton reactions 
are discussed in Section \ref{sec:elementary}, where 
the model calculations are compared to existing DLS data 
and predictions for the related
projects of the HADES Collaboration, 
$pp$ at 1.25 GeV, 2.2 GeV and 3.5 GeV, are presented.  
In Section \ref{dilepton_spectra}
dilepton spectra for C+C  
collisions are shown.
In Section \ref{sec:arkcl} we turn to predictions for the forthcoming 
analysis of dilepton production in Ar+KCl collisions .
Section \ref{sascha} is devoted to the study of the time evolution 
of the dilepton emission and its connection with the 
various density regimes experienced in the course of the heavy ion collision.
Summary and concluding remarks 
are finally given in Section
\ref{concl}.

\section{The model} \label{The model}
\subsection{Meson production in UrQMD}
In the UrQMD model the formation of light mesons
at low energies is modelled as a
multi-step process that proceeds via intermediate
heavy baryon and meson resonances and their subsequent decay
\cite{winckelmann95a}.
The resonance parameters (pole masses, widths
and branching ratios) are within the limits of \cite{PDG06}.
A comparison between the
exclusive and inclusive cross sections for the production of
neutral $\pi^0,\eta, \rho^0, \omega$ mesons in $pp$  reactions obtained 
within the UrQMD model and experimental data can be found in 
\cite{Bleicher:1999xi}. The resonant exclusive production of the 
$\rho^0$ meson, particularly important at low energies, will be discussed 
more in detail in Section \ref{discussion}.

\begin{figure}
\includegraphics[height=7.cm]{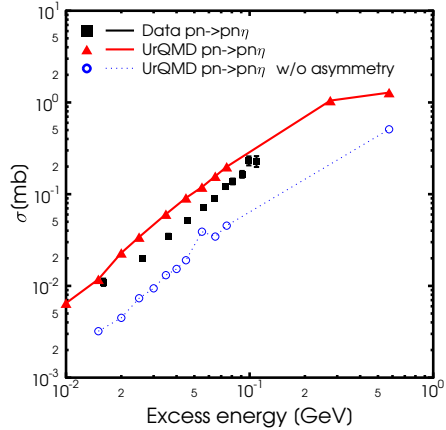}
\caption{(Color online) The $\eta$ production cross section from $pn$ reactions
 as a function of the excess energy. The UrQMD results 
obtained with the novel introduction of the isospin asymmetry in the $\eta$ 
production cross section (triangles) 
are compared to experimental data \cite{Calen:1998vh}. 
The circles refer to calculations which neglect such asymmetry and 
are shown for completeness.}
\label{fig:pn_eta}
\end{figure}

\begin{figure}
\includegraphics[height=7.cm]{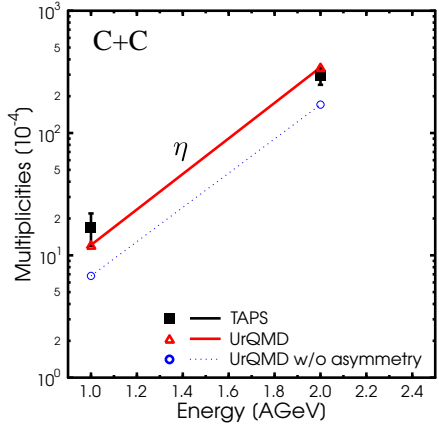} 
\caption{(Color online) Average $\eta$ multiplicity in C+C reactions at 
1 AGeV 
and 2 AGeV from UrQMD (triangle) in comparison to the values  
reported by the TAPS 
Collaboration \cite{Averbeck:1997ma}.
The circles refer to the standard calculations which neglect the 
isospin asymmetry in the $\eta$ 
production cross section and are shown for completeness.}
\label{fig:eta_multipli}
\end{figure}

In the analysis of dilepton spectra in 
nucleus-nucleus collisions performed with the UrQMD model in 
\cite{Ernst:1997yy}, 
the dilepton yield originating from the $\eta$ Dalitz decay 
was found about a factor two lower than in Ref. \cite{Bratkovskaya:1997mp} 
and Ref. \cite{Holzmann:1997mu}. In the latter, 
the $\eta$ channel had been determined 
from the measurements of the TAPS Collaboration.
As already anticipated in \cite{Ernst:1997yy}, the discrepancy could 
be attributable to the fact that the 
asymmetry in the $\eta$ production in $pp$ and $pn$ 
reactions ($\eta$ production cross sections in $pn$ reactions are 
about a factor five 
higher than in $pp$ reaction) had been neglected in the calculations. 
Such asymmetry has been  introduced for the present analysis (see Fig. \ref{fig:pn_eta}). 
The inclusion has been performed, 
as in \cite{Teis:1996kx}, at the level of the production cross section of the 
$N^{\star}(1535)$ resonance. 
For the C+C reactions  under study  the $\eta$ multiplicity 
obtained within the UrQMD model 
is now consistent with the value measured by the TAPS Collaboration  
\cite{Averbeck:1997ma}, as shown in Fig. \ref{fig:eta_multipli}.
The experimental constraint imposed by the TAPS measurements on the 
$\eta$ Dalitz contribution to the dilepton spectra in 
nucleus-nucleus collisions 
is thus respected by our calculations. Especially for C+C collisions 
at 2 AGeV, this is very important since, as we will see, the $\eta$ decay plays 
an important role in determining the spectra in the low mass region.

The energy dependence of the exclusive $pn\rightarrow pn\eta$ cross section as shown in 
Fig. \ref{fig:pn_eta} provides a reasonable description of the data, however a finer 
parametrization, as e.g. in 
Ref. \cite{Bratkovskaya:2007jk}, might be required in future studies of 
dilepton production in elementary  $pn$ reactions.  Especially for those 
cases where fixing the $\eta$ contribution with high precision is  
mandatory in order to achieve an unique interpretation of 
the experimental data in the low mass region a re-tuning is necessary. 
However, $pn$ reactions are
not the major subject of this work, and the new prescription used here for 
the treatment of $\eta$ production  provides 
sufficient robustness for the dilepton studies presented in the next sections.

\subsection{Dilepton radiation in UrQMD}
In UrQMD, dilepton pairs are generated from the mesonic Dalitz decays 
$\pi^0\rightarrow \gamma e^{+} e^{-}$, $\eta\rightarrow \gamma e^{+} e^{-}$, 
$\eta'\rightarrow \gamma e^{+} e^{-}$ and 
$\omega\rightarrow \pi^0 e^{+} e^{-}$, 
the direct decay of the $\rho$, $\omega$ and $\phi$ vector mesons and the 
Dalitz decay of the $\Delta$ resonance.

Decays of the form, with $P$ being a pseudoscalar meson and $V$ a vector meson,
\begin{equation} 
P \rightarrow \gamma  e^{+} e^{-},  V \rightarrow P e^{+} e^{-}
\end{equation}
can be decomposed into the corresponding 
decays into a virtual photon $\gamma^{\star}$, 
$P \rightarrow \gamma \gamma^{\star}$, $ V \rightarrow P \gamma^{\star}$,
and the subsequent decay of the photon via electromagnetic
conversion, $\gamma^{\star} \rightarrow e^{+} e^{-}$ 
\cite{Landsberg:1986fd,Koch:1992sk,Faessler:1999de}:
\beqa
\frac{d\Gamma_{P \rightarrow \gamma  e^{+} e^{-}}}{dM^2}=
\Gamma_{P \rightarrow \gamma \gamma^{\star}}\,\frac{1}{\pi M^4}\,
M\Gamma_{\gamma^{\star} \rightarrow e^{+} e^{-}}~,\label{pseudo_dalitz_0}\\
\frac{d\Gamma_{V \rightarrow P e^{+} e^{-}}}{dM^2}=
\Gamma_{V \rightarrow P \gamma^{\star}}\,\frac{1}{\pi M^4}\,
M\Gamma_{\gamma^{\star} \rightarrow e^{+} e^{-}}~,
\eeqa
where $M$ is the mass of the virtual photon or, equivalently, the 
invariant mass of the lepton pair. The internal conversion probability of the 
photon is given by:
\beq
M\Gamma_{\gamma^{\star} \rightarrow e^{+} e^{-}}=
\frac{\alpha}{3}\,M^2\,\sqrt{1 - \frac{4m_{e}^{2}}{M^{2}}} \left(1 + \frac{2
  m_{e}^{2}}{M^{2}} \right)
\eeq
with $m_e$ being the electron mass.
The widths $\Gamma_{P \rightarrow \gamma \gamma^{\star}}$ and  
$\Gamma_{V \rightarrow P \gamma^{\star}}$ 
can be related to the corresponding radiative widths 
$\Gamma_{P \rightarrow 2 \gamma}$ and 
$\Gamma_{V \rightarrow P \gamma}$: 
\begin{eqnarray}
\Gamma_{P \rightarrow \gamma \gamma^{\star}}&=&2\,
\Gamma_{P \rightarrow 2 \gamma}\,\left(1 - \frac{M^{2}}{m_{P}^{2}} \right)^{3}
|F_{P\gamma\gamma^{\star}} (M^{2})|^{2} , \label{pseudo_dalitz}\\
\Gamma_{V \rightarrow P \gamma^{\star}}&=&
\Gamma_{V \rightarrow P \gamma}\,\left[ 
\left(1 + \frac{M^{2}}{m_{V}^{2}-m_{P}^{2}}\right)^{2}  - \left(\frac{2m_{V}M}{m_{V}^{2}-m_{P}^{2}}\right)^{2} \right]^{3/2} |F_{V P \gamma^{\star}}(M^{2})|^{2},
\end{eqnarray}
where $m_{P}$ and $m_{V}$ are the masses of the pseudoscalar and vector meson 
respectively and $F_{P\gamma\gamma^{\star}} (M^{2})$, $F_{V P \gamma^{\star}}(M^{2})$ 
denote the form factors with  
$F_{P\gamma\gamma^{\star}} (0)=F_{VP \gamma^{\star}}(0)=1$. 
The factor 2 in (\ref{pseudo_dalitz}) occurs due to the identity of the two 
photons in the $P \rightarrow 2 \gamma$ decay. The form factors can be 
obtained from the vector meson dominance
model (VMD). In the present calculations 
the following parametrisations are employed
\cite{Landsberg:1986fd,Li:1996mi}:
\beqa \label{dalitz}
F_{\pi^0}(M^2)&=&1+b_{\pi^0} M^2,\nonumber\\
F_{\eta}(M^2)&=&\left(1-\frac{M^2}{\Lambda_\eta^2}\right)^{-1},\nonumber\\
\left|F_\omega(M^2)\right|^2&=&
\frac{\Lambda_\omega^2(\Lambda_\omega^2+\gamma_\omega^2)}
{(\Lambda_\omega^2-M^2)^2+\Lambda_\omega^2\gamma_\omega^2},\nonumber\\
\left|F_{\eta'}(M^2)\right|^2&=&
\frac{\Lambda_{\eta'}^2(\Lambda_{\eta'}^2+\gamma_{\eta'}^2)}
{(\Lambda_{\eta'}^2-M^2)^2+\Lambda_{\eta'}^2\gamma_{\eta'}^2}
\label{eq_VMDform}
\eeqa
with $b_{\pi^0}=5.5\GeV^{-2}$, $\Lambda_\eta=0.72\GeV$,
$\Lambda_\omega=0.65\GeV$,
$\gamma_\omega=0.04\GeV$,
$\Lambda_{\eta'}=0.76\GeV$ and
$\gamma_\eta'=0.10\GeV$. In (\ref{eq_VMDform}) 
the abbreviations $F_P$ and $F_V$ 
have been used to denote respectively 
$F_{P\gamma\gamma^{\star}}$ and $F_{VP \gamma^{\star}}$.

The width for the direct decay of a vector meson 
$V = \rho^{0},~ \omega,~ \phi$ to a dilepton pair  varies with
the dilepton mass like $M^{-3}$ according to \cite{Li:1996mi}:
\begin{equation} \label{direct}
\Gamma_{V \rightarrow e^{+} e^{-}}(M) = \frac{\Gamma_{V \rightarrow e^{+}
e^{-}}(m_{V})}{m_{V}} \frac{m_{V}^{4}}{M^{3}} \sqrt{1 - \frac{4m_{e}^{2}}{M^{2}}}
\left(1 +  \frac{2m_{e}^{2}}{M^{2}} \right)
\end{equation}
with $\Gamma_{V \rightarrow e^{+} e^{-}}(m_{V})$ being the partial decay 
width at the
meson pole mass. 

The decomposition of the $\Delta \rightarrow N e^{+} e^{-}$ decay  into the 
$\Delta \rightarrow N \gamma^{\star}$ decay and subsequent conversion of the 
photon leads to the following expression for the differential decay width:
\beq \label{delta}
\frac{d\Gamma_{\Delta \rightarrow N  e^{+} e^{-}}}{dM^2}=
\frac{\alpha}{3 \pi M^2}\Gamma_{\Delta \rightarrow N \gamma^{\star}}~.
\eeq
Here the electron mass has been neglected.
The decay width into a massive photon reads~\cite{Wolf:1990ur}:
\begin{eqnarray} \label{decay_Delta1}
\Gamma_{\Delta \rightarrow N \gamma^{\star}}(M_{\Delta}, M) &=& 
\frac{\lambda^{1/2}(M^{2}, m_{N}^{2}, M_{\Delta}^{2})}{16 \pi
M_{\Delta}^{2}} m_{N} \nonumber \\
 & & \times~~ [2 \mathcal{M}_{t}(M, M_{\Delta}) + \mathcal{M}_{l}(M,
M_{\Delta})]~,
\end{eqnarray}
where the kinematic function $\lambda$ is defined by 
$\lambda(m_{A}^{2}, m_{1}^{2}, m_{2}^{2})
 = (m_{A}^{2}-(m_{1}+m_{2})^{2})(m_{A}^{2}-(m_{1}-m_{2})^{2})$ and 
$M_{\Delta}$ is the resonance running mass.  The matrix
elements $\mathcal{M}_{t}$ and $\mathcal{M}_{l}$  are taken 
from \cite{Wolf:1990ur}. The coupling constant $g$ appearing in the expression for $\mathcal{M}_{t}$ and $\mathcal{M}_{l}$ has been chosen as $g=5.44$, in order to reproduce the value of the radiative decay width, as done e.g. in \cite{Bratkovskaya:1999mr}.

\subsection{Shining method}
The "shining" method (also called time integration method) was 
introduced in \cite{Heinz:1991fn} and
\cite{Li:1994cj} and assumes that a resonance can continuously 
emit dileptons over its whole lifetime. 
The dilepton yield is obtained by integration 
of the dilepton emission rate over time, taking the 
collisional broadening of each individual parent resonance into account:
\begin{equation}\label{shining}
\frac{dN_{e^{+} e^{-}}}{dM} =\frac{\Delta N_{e^{+} e^{-}}}{\Delta M}= \sum _{j=1}^{N _{\Delta M}} \int _{t_i^j} ^{t_f^j} \frac{dt}{\gamma } \frac{\Gamma_{e^{+} e^{-}}(M)}{\Delta M}
\end{equation}
Here  $\Gamma _{e^+e^-}(M)$ is the electromagnetic decay 
width of the considered resonance defined in 
(\ref{direct}--\ref{decay_Delta1}) and $t=t_i$ ($t_f$) 
the time at which the resonance appeared in (disappeared from) the system. 
For the calculations applying the "shining" method 
the whole time evolution of the collision is reconstructed. 
Each resonance is followed 
from the production time $t_i$ to a final time $t_f$ 
at which the resonance decays or is 
reabsorbed.
We implement the shining method for the short-lived vector 
mesons $\rho$ and $\omega$ and 
the baryonic resonance $\Delta$. 

In an alternative method, dileptons have been extracted at the point of decay 
of the resonances \cite{Schumacher:2006wc}. 
The dilepton yield is calculated at the decay vertex 
from the branching ratio. Thus, in this method the contribution 
to the dilepton yield of the reabsorbed resonances is neglected. As 
shown in \cite{Vogel:2007yu} this contribution is however small.
A comparison between the two methods is realised in the next section.

\section{Elementary reactions}
\label{sec:elementary}

\subsection{Comparison to DLS measurements}

Before addressing heavy ion collisions we consider dilepton production in elementary reactions. The 
latter are very important to gain a better understanding of the various processes 
contributing to the dilepton production and of their relative weights. 
In the energy range of interest for this work there exist measurements from the DLS \cite{Wilson:1997sr} 
and HADES Collaboration.

\begin{figure}[h]
\includegraphics[height=16.0cm]{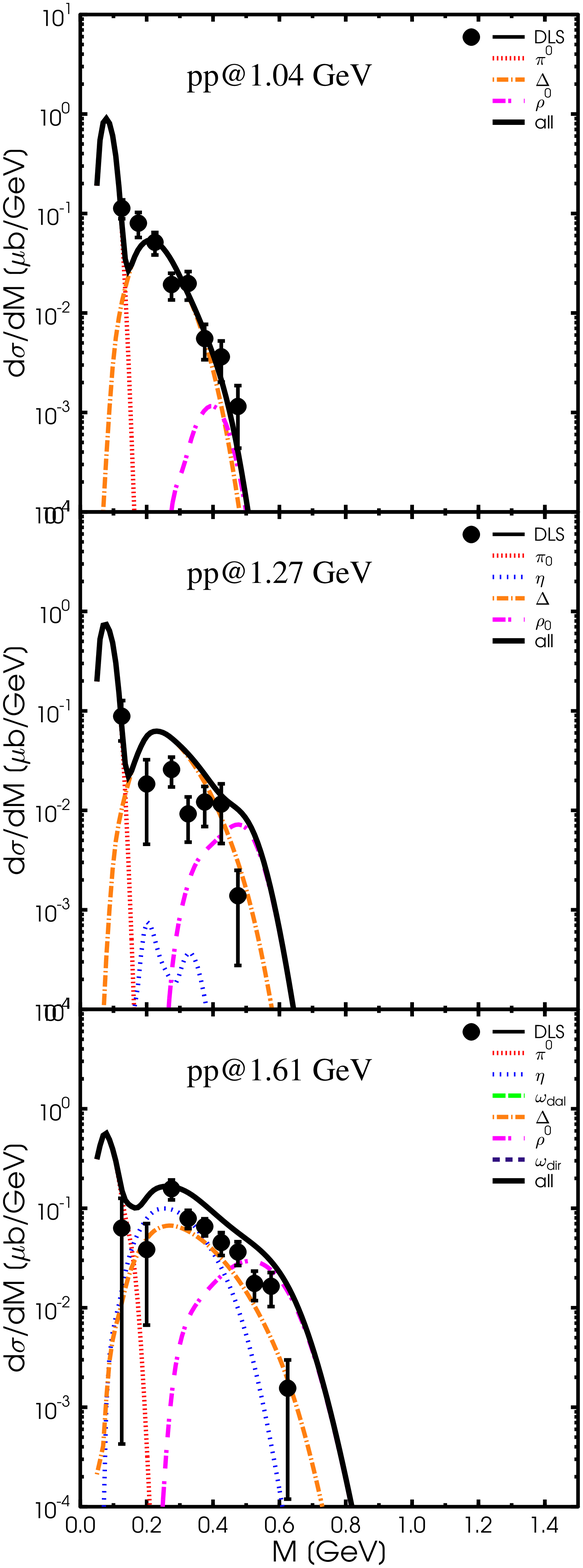}
\hspace*{-.4cm}
\includegraphics[height=16.0cm]{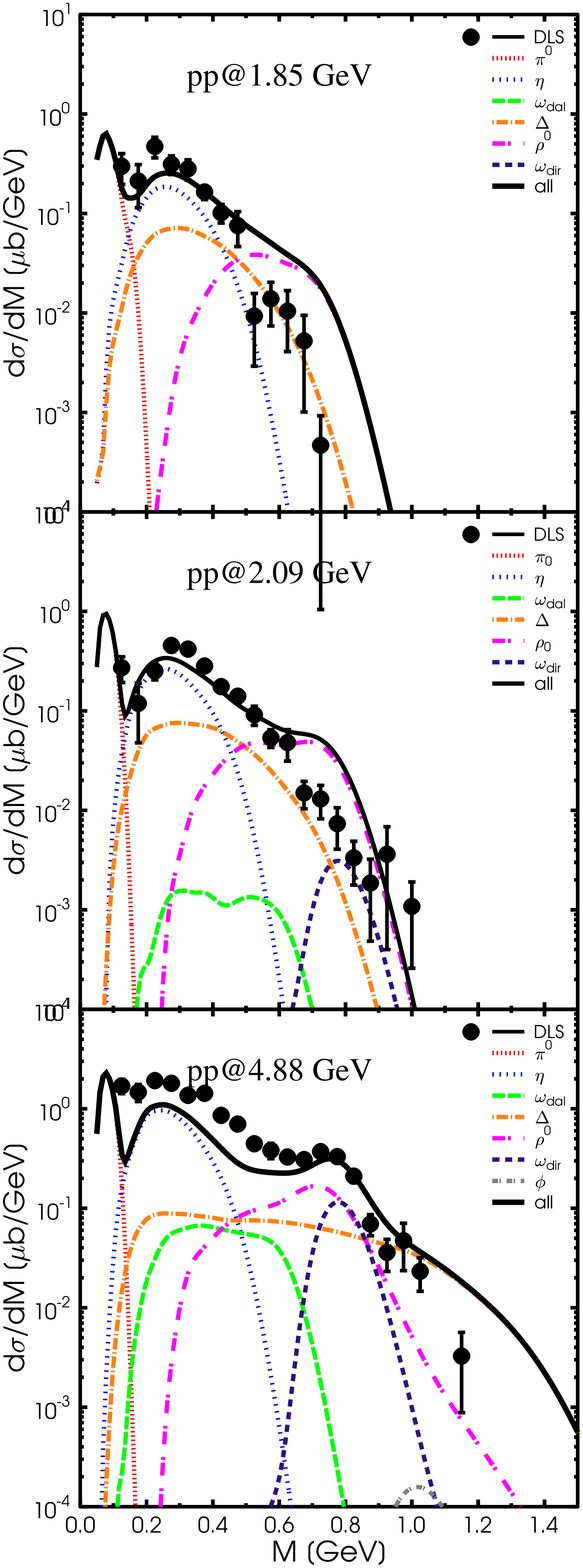}
\caption{(Color online) UrQMD model calculations
for dilepton spectra from
$pp$ reactions at 1.04, 1.27, 1.61, 1.85 , 2.09 and 4.88 GeV 
in comparison to the DLS data \cite{Wilson:1997sr}, 
including the DLS acceptance filter and mass resolution. 
The different color 
lines display individual channels in the transport calculation, 
as indicated in the legend. \label{fig:dls_pp}}
\end{figure}

Differential dilepton cross sections have been calculated with the present model 
for $pp$ reactions  at beam energies of 1.04, 1.27, 1.61, 1.85 , 2.09 and 4.88 GeV.
The results are presented in Fig. \ref{fig:dls_pp} in comparison to the DLS 
data \cite{Wilson:1997sr}. 
In order to perform the comparison, the DLS acceptance filter and mass 
resolution have been included. For collisions at 1.04--2.09 GeV 
the agreement with the available data is generally
reasonable in the region $M\leq 0.45$ GeV, where the $\pi^0$,  $\Delta$  and $\eta$  Dalitz 
decays dominate, a systematic overestimation of the data is observed at higher masses. 
Especially at 2.09 GeV a clear overestimation of the 
dilepton cross section 
around the vector meson peak is present, a result which is 
analogous to the findings of Ref. \cite{Bratkovskaya:2007jk}. This might be due to an insufficient modelling 
of the  production rate of high mass resonances in $pp\rightarrow pN^*$, $pp\rightarrow p\Delta^*$ collisions. 
We investigate this effect more in detail in Section \ref{discussion}.
At bombarding energy of 4.88 GeV an inversion of this trend 
is observed and data are underestimated by the model calculations  
in the low invariant mass region but good described in the vector meson region. 
This is not a contradiction. The main difference lies in the fact that at 4.88 GeV 
the exclusive production of the $\rho$ meson does not 
affect significantly the inclusive production. The latter, on the other side, determines the $\rho$ meson 
yields in the  reactions at 1.04--2.9 GeV.

\subsection{Predictions for HADES}
\begin{figure}[htb!]
\includegraphics[width=.37\textwidth]{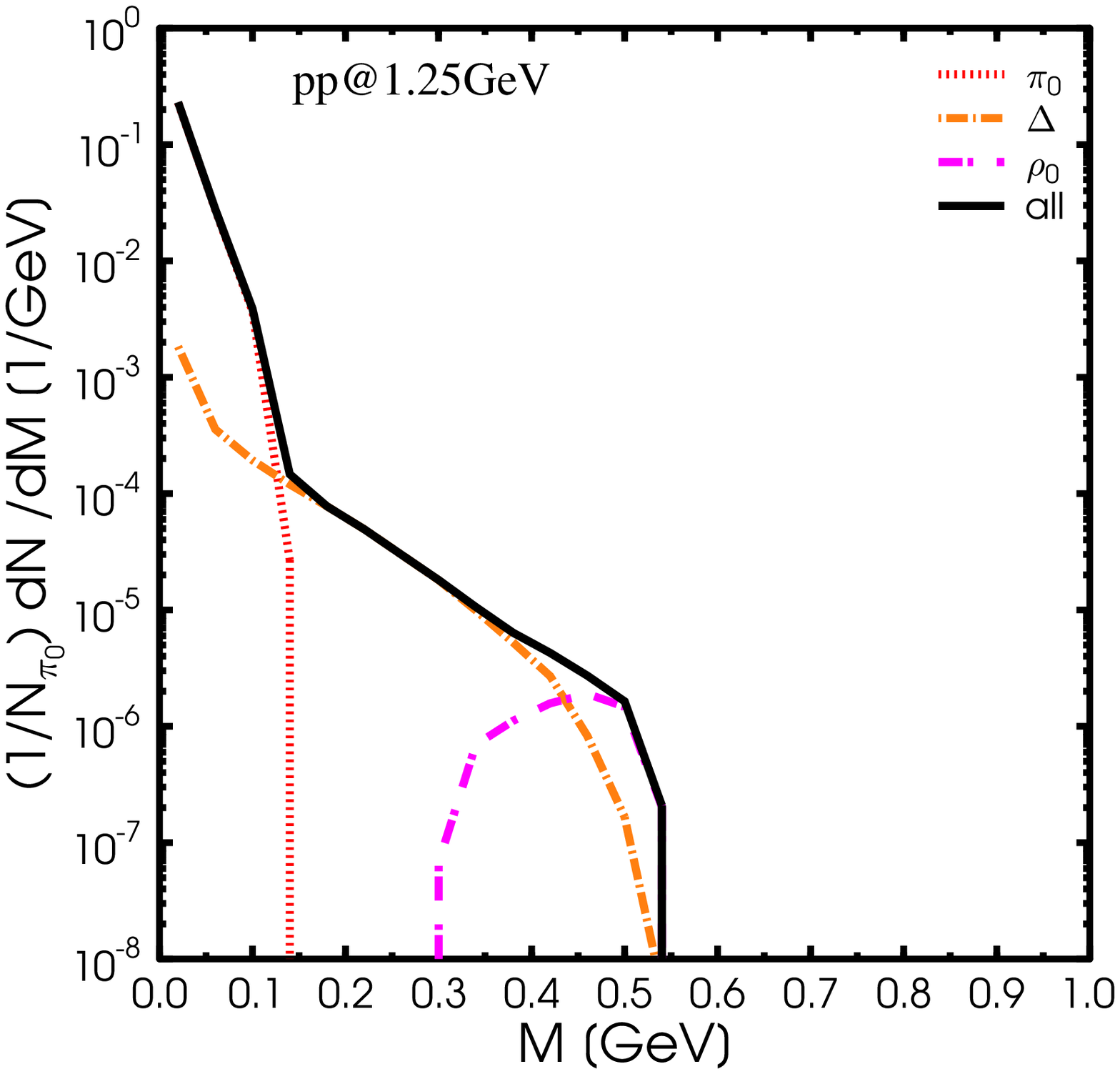}
\hspace*{-1.2cm}
\includegraphics[width=.37\textwidth]{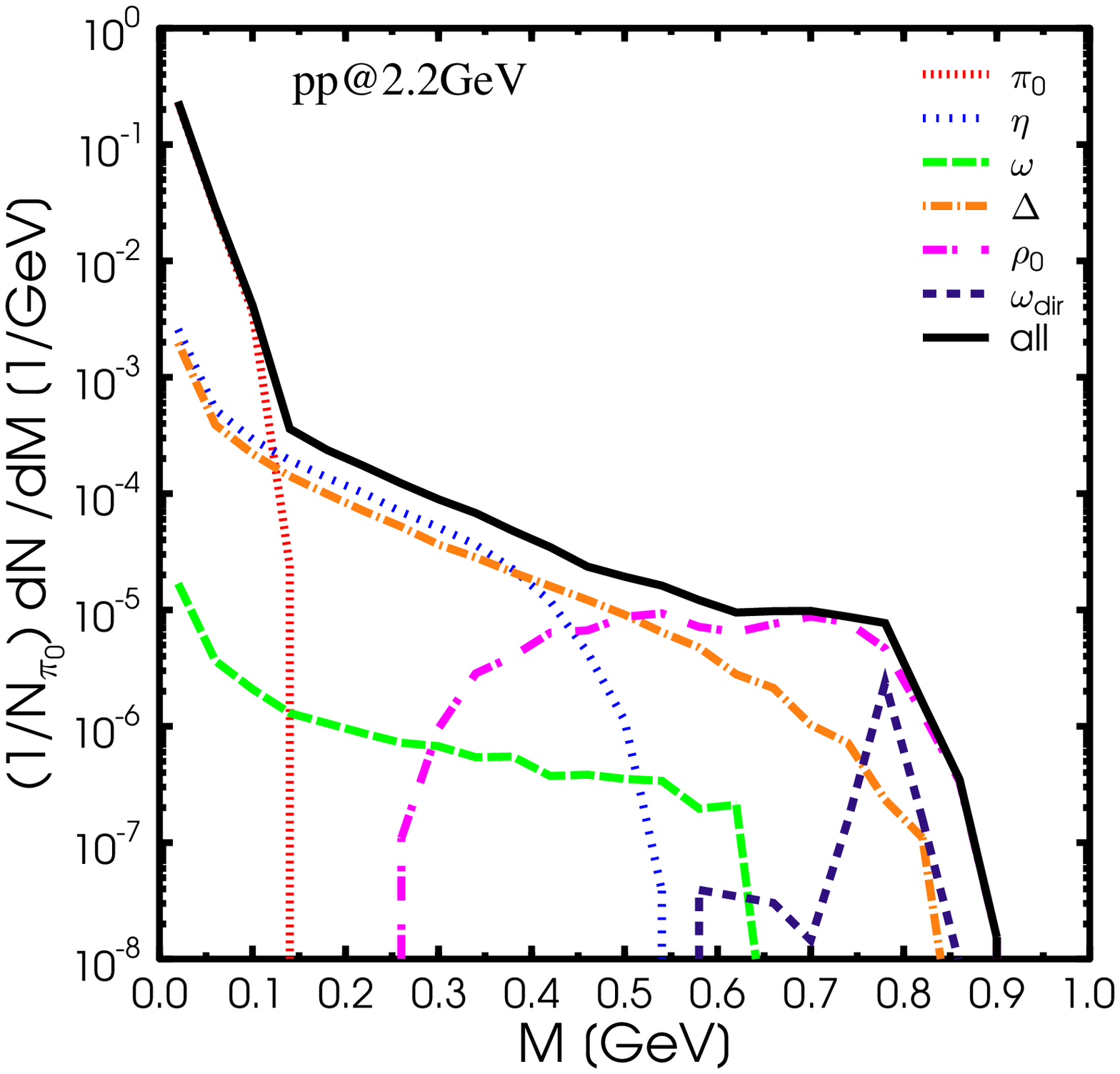}
\hspace*{-1.2cm}
\includegraphics[width=.37\textwidth]{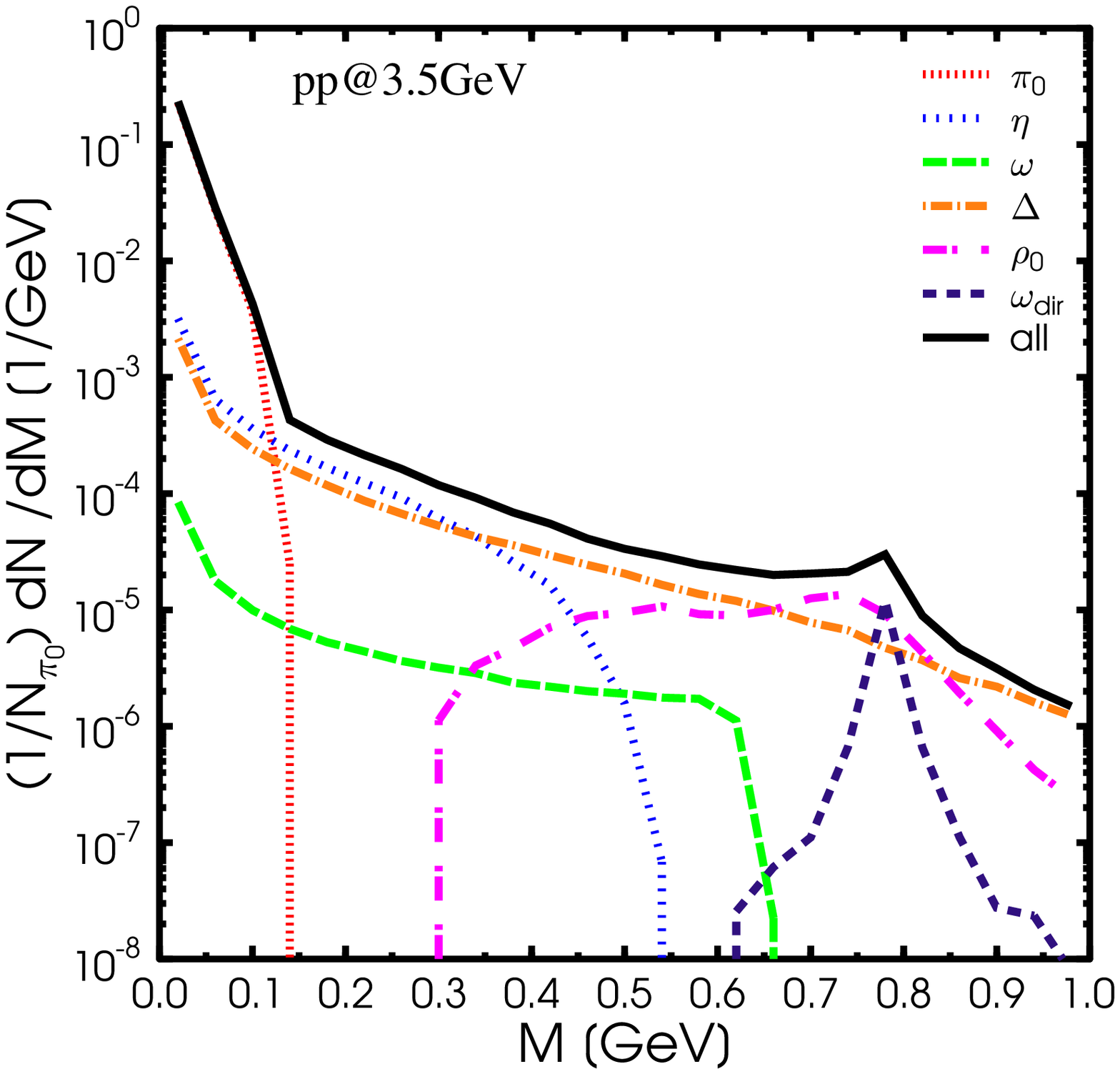}
\caption{(Color online) UrQMD model calculations
for dilepton spectra from
p+p collisions at beam energies of 1.25~GeV (left panel), 2.2~GeV (middle panel) and 3.5~GeV (right part). The different color 
lines display individual channels from the transport calculation, 
as indicated in the legend. \label{fig:pp_mass_unfil}}
\end{figure}

The HADES physic program includes measurement of $pp$ reactions at 1.25~GeV, 2.2~GeV and 3.5~GeV which we want to discuss here. 
In Fig. \ref{fig:pp_mass_unfil}, UrQMD calculations for the three energies are presented. 
The beam energy $E = 1.25$~GeV is below the $pp \rightarrow pp\eta$ threshold and is therefore optimal for studying  the contribution from $\Delta$ Dalitz. For $M > 0.45$~GeV 
a noticeable contribution from $\rho ^0 \rightarrow e^+e^-$ is visible. This result differs from other calculations \cite{Bratkovskaya:2007jk}, where the contribution from the direct decay of the $\rho $ meson is not seen at the lowest energy. 
This is due to the omission of an explicit treatment of $\rho$ meson production via resonant mechanism in \cite{Bratkovskaya:2007jk}, 
where a simplified parametrization of the  $pp\rightarrow \rho X$ (vacuum) cross 
section of the form $\mbox{$\sigma(pp\rightarrow \rho X)\sim\int 2.2 ~(\frac{s}{s_0(M)}-1)^{1.47}~(\frac{s}{s_0(M)})^{-1.1}~A(M)~dM$}$  has been employed. Here $A(M)$ denotes the meson spectral function and the integration is performed within the appropriate kinematical limits.
Close to the physical threshold for 
$\rho$ meson production, $\sqrt{s}_{\rm th}=2m_N+2m_\pi$, such omission results in smaller values of 
the cross section than those of this work (see Section \ref{discussion}) and of other resonance model based approaches (see e.g. \cite{Teis:1996kx,Faessler:2000md,Shekhter:2003xd}).
In our model, 
this contribution  arises naturally due to the possibility for baryonic resonances to decay into $\rho$. 
At rather low energies, this leads to the emission of a $\rho$ meson with a mass distribution strongly biased by energy constraints.
Here, the $\rho$ mesons originates 
in particular from the decay of the $N^*(1520)$ resonance. 
For this chain the threshold is only $M = 2m_{\pi}$ and not $m_\rho^{\rm pole}$. Early investigations on the role of the $N^*(1520)$ resonance for subthreshold 
$\rho$ meson production were performed in Refs.~\cite{Bratkovskaya:1999mr,Bratkovskaya:1998pr}.

For higher beam energies all decays are possible as for the nucleus nucleus system. 
Both for 2.2~GeV and 3.5~GeV the dilepton spectra in the lower mass regime are dominated by the 
long-lived resonances and the $\Delta$ resonance.   
For higher masses the direct decay of the $\rho$ meson becomes more important 
and the double peak shape of the $e^+e^-$-pairs originating from $\rho$ is visible. At a beam energy of 3.5~GeV the contribution from the direct $\omega$ decay leads to a visible peak in the dilepton spectrum at $M \approx 0.8$  GeV.

\section{Dilepton yields in C+C collisions} \label{dilepton_spectra}
In this Section we present calculations for dilepton spectra 
in minimum bias C+C reactions at 1.0 AGeV and 2.0 AGeV and compare them to the data resulting from the 
measurements performed by the HADES Collaboration \cite{Agakichiev:2006tg,Agakishiev:2007ts}.
In order to make the comparison with the experimental data, 
the filter function provided by the HADES Collaboration has been implemented \cite{Agakichiev:2006tg,Agakishiev:2007ts}. 
In agreement with the treatment of the experimental data, 
dilepton events with opening angle $\Theta_{e^{+} e^{-}} \leq 9^{\circ} $ have been rejected and 
the spectra have been normalised to the 
mean $\pi^0$ multiplicity.

\begin{figure}[hbt]
\includegraphics[height=8cm]{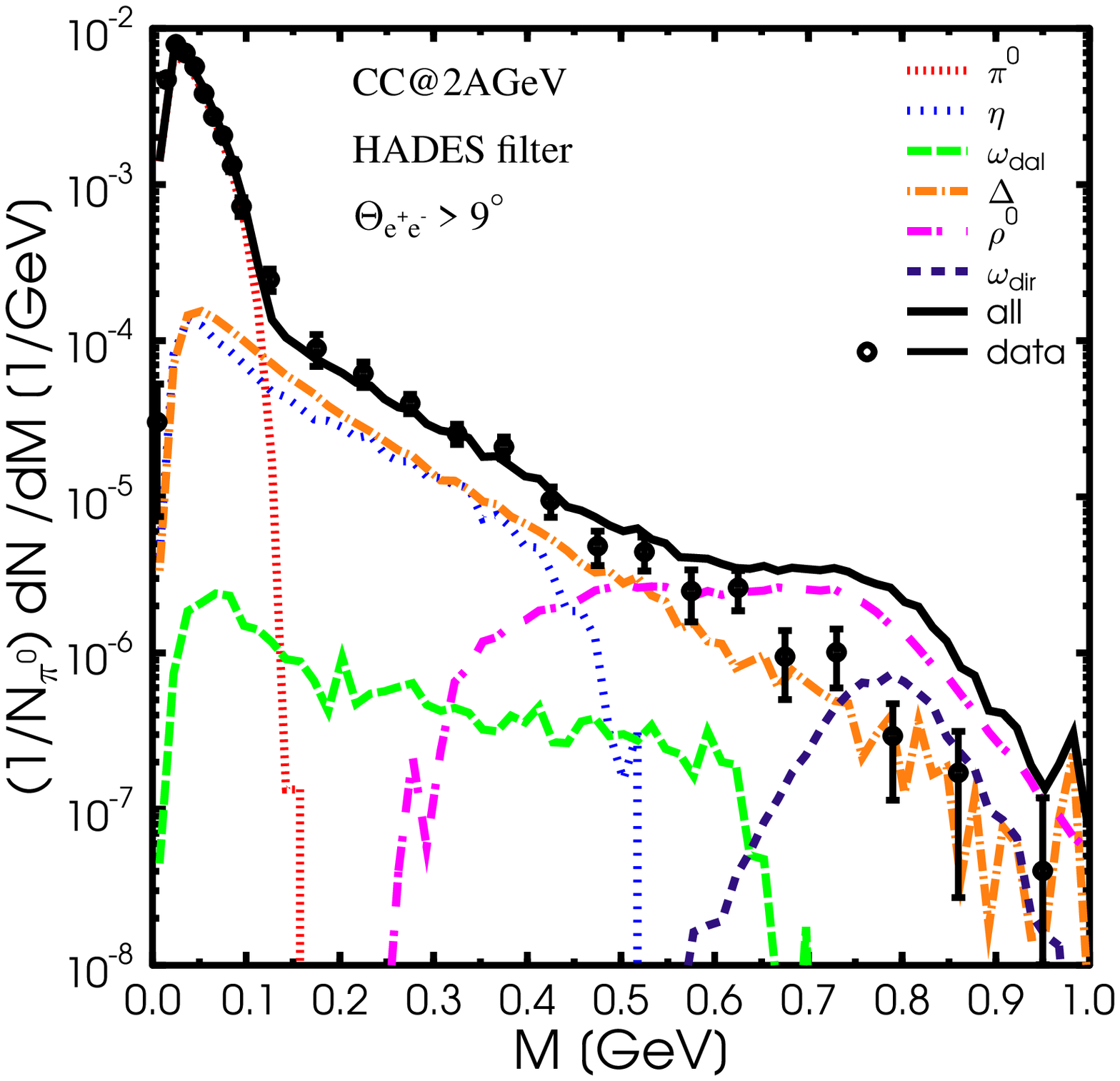}
\hspace*{-1cm}
\includegraphics[height=8cm]{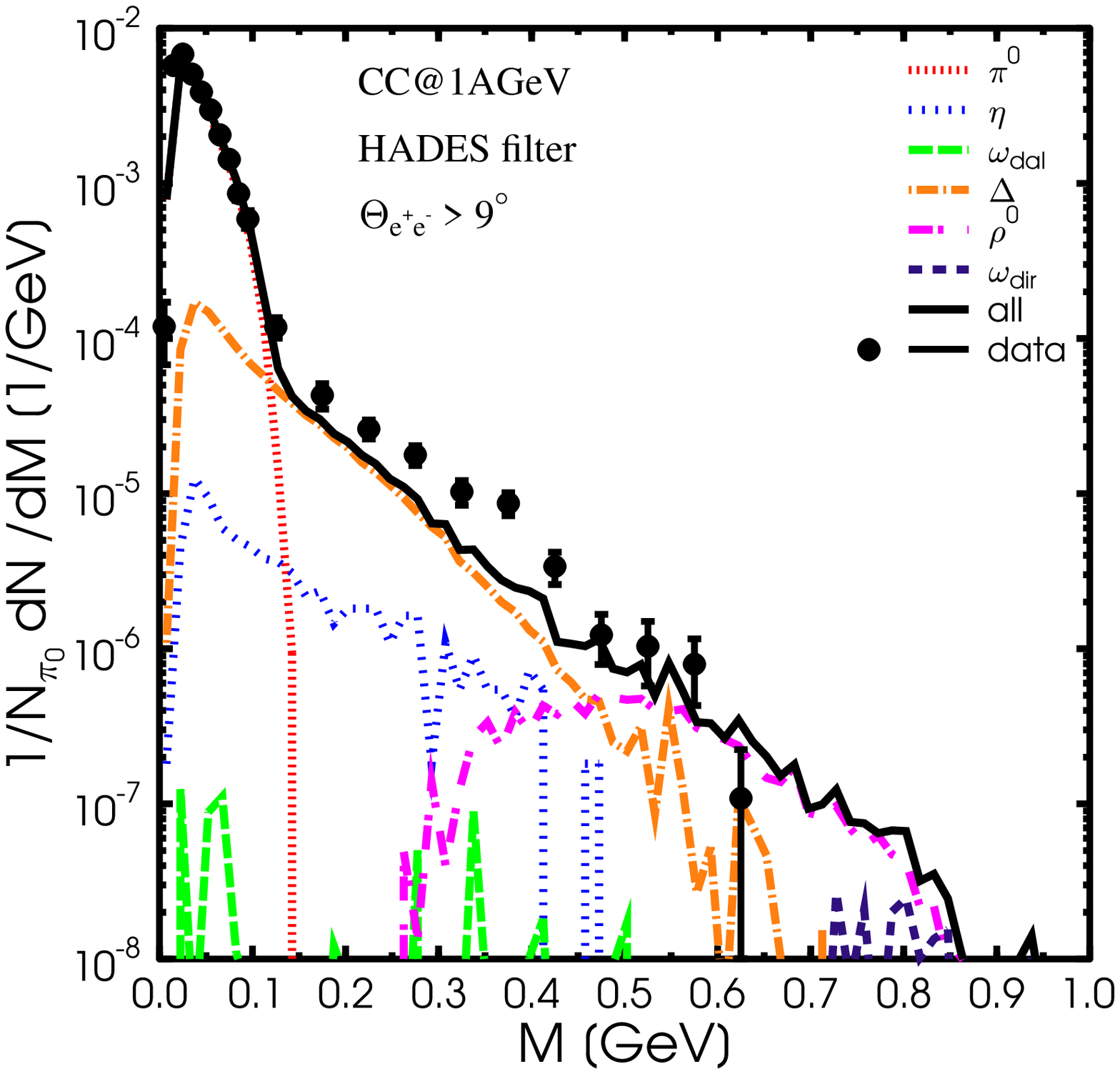}
\caption{(Color online) UrQMD model calculations
for dilepton spectra from
C+C collisions at beam energies of  2~AGeV (left) and  1~AGeV (right) in comparison to 
HADES data \cite{Agakichiev:2006tg}. The different color 
lines display individual channels in the transport calculation, 
as indicated in the legend.
\label{filter_cc2}}
\end{figure}

We first discuss the results obtained applying the ``shining'' method 
for the extraction of the dilepton yield 
and address Fig. \ref{filter_cc2}, where the 
contributions to the spectra of  the different channels are additionally explicitly shown.
Both spectra are dominated by the $\pi^0$ decay for invariant masses 
$M \leq m_{\pi}$. 

In the case of C+C at  2 AGeV the $\eta$ and $\Delta$ Dalitz decays dominate for $m_{\pi}\leq  M\leq 0.5$ GeV with comparable magnitude. 
The present result for the $\Delta$ Dalitz contribution to the spectra is quantitatively similar to 
the result of Ref. \cite{Bratkovskaya:2007jk} , whereas in \cite{Thomere:2007cj} and \cite{Santini:2008pk} 
a smaller contribution was found. 
The direct decay of the $\rho$ meson start to play a 
sizable role for $M\geq 0.5$ GeV. 
Due to the rapid decrease of the  $\Delta$ 
Dalitz contribution, the  relative importance of the $\rho$ meson direct 
decay channel grows with increasing invariant mass, 
  from being at first comparable to the  $\Delta$ Dalitz to becoming the dominant contribution 
  in the region of the vector meson peak.
The low invariant mass region 
of the spectrum ($M < 0.5$ GeV)  is successfully 
described by the UrQMD calculations. 
However, an overestimation of the data is 
observed at higher masses.
A qualitatively analogous result has been 
found in the analysis of \cite{Bratkovskaya:2007jk}, were 
the ``vacuum'' calculation for C+C at 2 AGeV resulted in an overestimation of the data in the 
region of the vector 
meson peak. However, the enhancement being  more localised 
around the peak than in our case and about a factor 1.5 lower 
at $M\sim m_{peak}$. The difference lies
in the contribution originating from the direct $\rho$ meson decay, 
suggesting a probably different value of $\rho$ meson multiplicity. 

\begin{figure}[hbt]
\includegraphics[height=8cm]{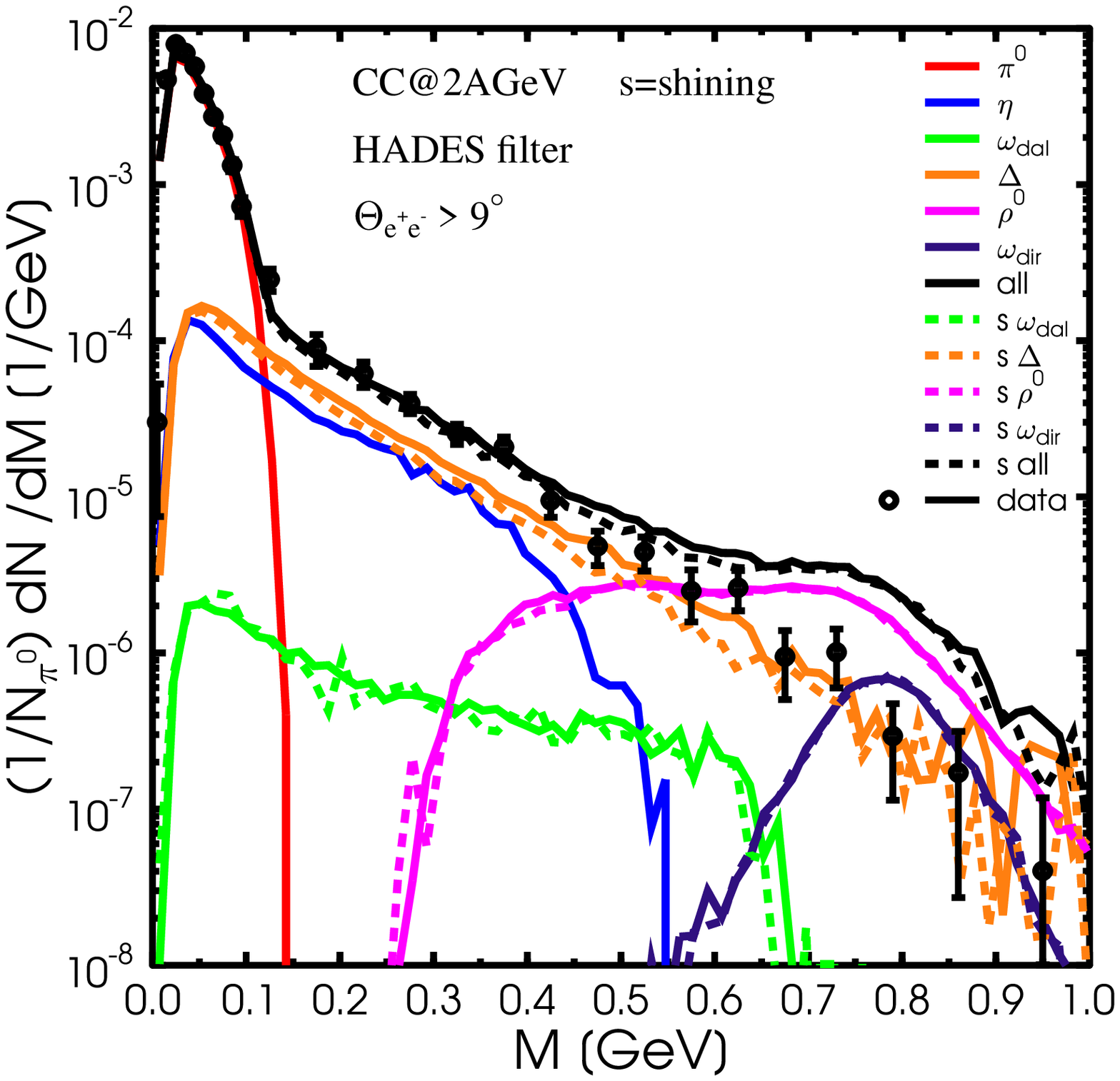}
\hspace*{-1.cm}
\includegraphics[height=8cm]{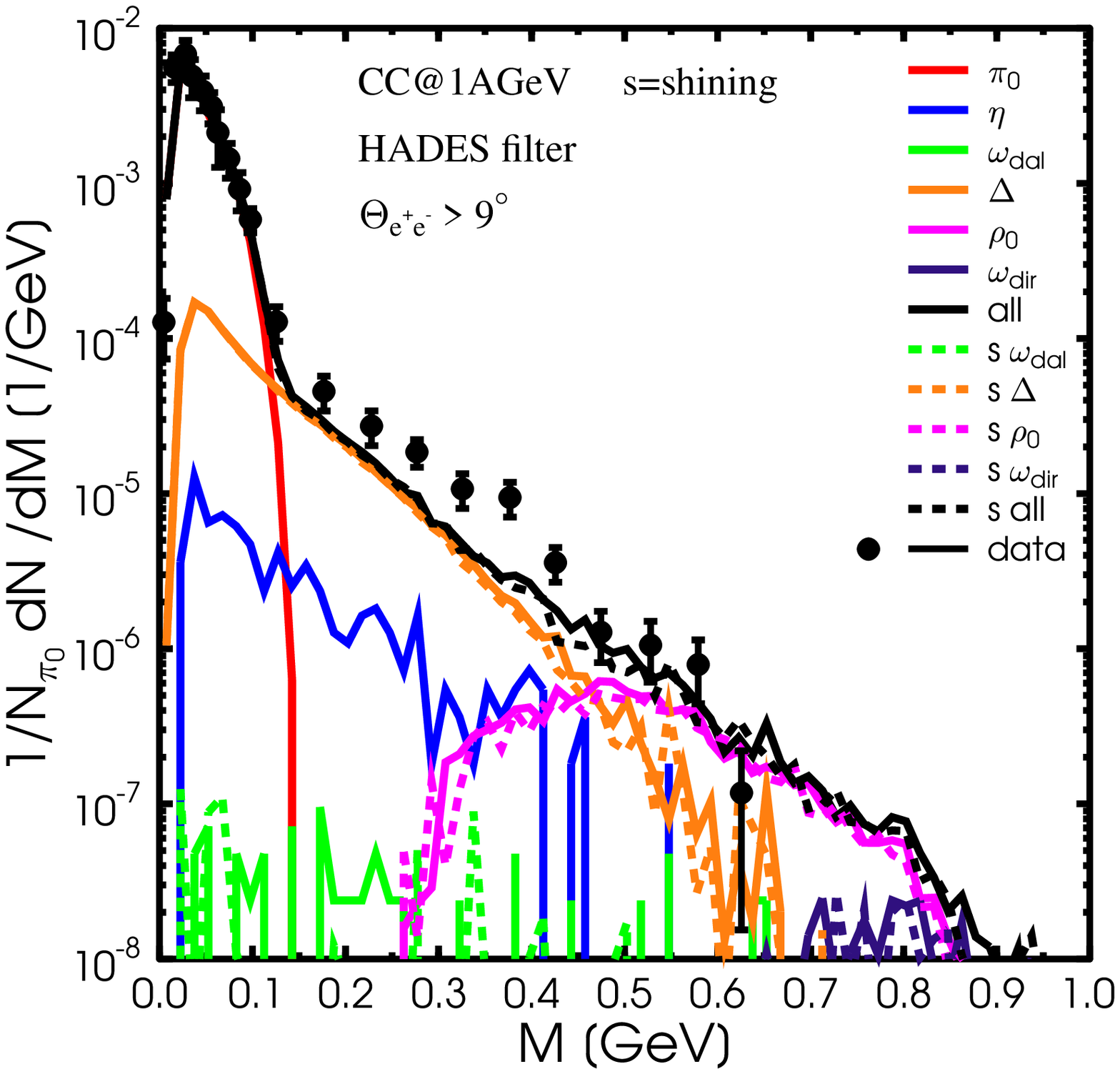}
\caption{(Color online) UrQMD model calculations
for dilepton spectra from
C+C collisions at beam energies of 2~AGeV (left) and 1~AGeV (right) in comparison to 
HADES data \cite{Agakichiev:2006tg}. 
The full lines correspond to determination of the dilepton 
yield at the decay vertex of the parent particle. 
The dashed lines 
correspond to the dilepton yield resulting from the 
application of the shining method. 
The different color 
lines display individual channels in the transport calculation, 
as indicated in the legend, with s indicating the shining method. \label{filter_cc2_shining}}
\end{figure}

The spectrum obtained assuming that dileptons are emitted at 
the decay vertex of the parent resonance 
is shown 
in Fig.~\ref{filter_cc2_shining}  and compared to the result of 
Fig.~\ref{filter_cc2}. The two results present no 
sizable differences, indicating that the methods to extract dileptons 
are essentially equivalent 
when looking at time integrated yields at low energies. 
The reason for that lies in the smallness 
of the yield originating from reabsorbed resonances if compared to 
the emission from decaying resonances \cite{Vogel:2007yu}.  
The effect of absorption processes on the dilepton spectrum is 
analysed in the Section \ref{sascha}.

Unfortunately no inclusive data on $\rho$ meson production cross section are 
available at the energy interesting for this work. 
Whether the here observed overestimation of the HADES data is due to 
an overestimation of the $\rho$ meson multiplicities from the nucleon-nucleon collisions or 
to a more complex in-medium mechanism, or both, 
cannot be decided on the basis of this experimental data. 
A comparison of the mass differential dilepton cross section for $pp$ reactions to existing 
DLS data has been performed and discussed in the previous section. 
The analysis suggested that the meson multiplicity might be indeed 
slightly overshot. 
Due to the low resolution of the DLS data, it is for the moment not possible to make exact 
quantitative conclusions. 
Nevertheless, an attempt to estimate possible model 
incertainties and their consequences is made in the next section.
In this respect, the forthcoming HADES data on 
dilepton production in elementary 
reactions will be extremely helpful to indirectly constrain vector meson multiplicities.

At 1 AGeV a systematic underestimation of the data is 
observed in the mass region $0.2 < M < 0.4$ GeV with a maximum discrepancy at 
$M\approx 0.38$ GeV. 
The result is qualitatively in line with previous investigations of dilepton 
production in 1 AGeV nucleus-nucleus collisions which link back in time to 
the DLS era \cite{Bratkovskaya:1997mp,Ernst:1997yy,Shekhter:2003xd}. 
Quantitatively, however, the discrepancy between 
the theoretical and experimental spectra spans here between a factor 
1.5 and 2 from $M=0.225$ GeV to $M=0.325$ and is then at most of a factor 3 
at $M=0.375$ GeV, whereas discrepancies of a factor four had emerged from 
the studies performed in the nineties \cite{Bratkovskaya:1997mp,Ernst:1997yy}.
Enhanced bremsstrahlung cross sections in line with one boson exchange calculations 
by Kaptari and K\"{a}mpfer \cite{Kaptari:2005qz} have been recently proposed as possible explanation of the DLS puzzle \cite{Bratkovskaya:2007jk}. 
The issue is however quite controversial. For $pn$ reactions the 
cross sections of Ref. \cite{Kaptari:2005qz} 
differ up to a factor four from previous calculations \cite{Schafer:1989dm,Shyam:2003cn}. In  \cite{Kaptari:2005qz} and \cite{Shyam:2003cn} 
the same couplings have been used, but differences can originate due to a different prescription used by the groups to restore gauge invariance  
in the effective theory. Since the way this restoration can be achieved is not unique, there are no straight 
arguments which favor one calculation over the other.  
To investigate this discrepancy, dilepton production in nucleon-nucleon collisions has 
been recently revisited within a fully relativistic and gauge invariant framework \cite{Shyam:2008rx}. 
For the various contributions analyzed --$pp$ bremsstrahlung, $pn$ bremsstrahlung, 
as well as contributions with the $\Delta$ isobar intermediate state-- the authors 
of \cite{Shyam:2008rx} found cross sections  smaller than those 
in Ref. \cite{Kaptari:2005qz}. In $pn$ collisions at beam energies of $1.04$ and 
$2.09$ GeV, in particular, differences in the bremsstrahlung contribution by factors 
ranging between 2 and 3 were foud. 
Future HADES measurements of dilepton spectra in elementary, especially $pn$, 
collisions will help to shed light into this new puzzle.

\subsection{Discussion \label{discussion}}
\begin{figure}[hbt]
\includegraphics[height=8cm]{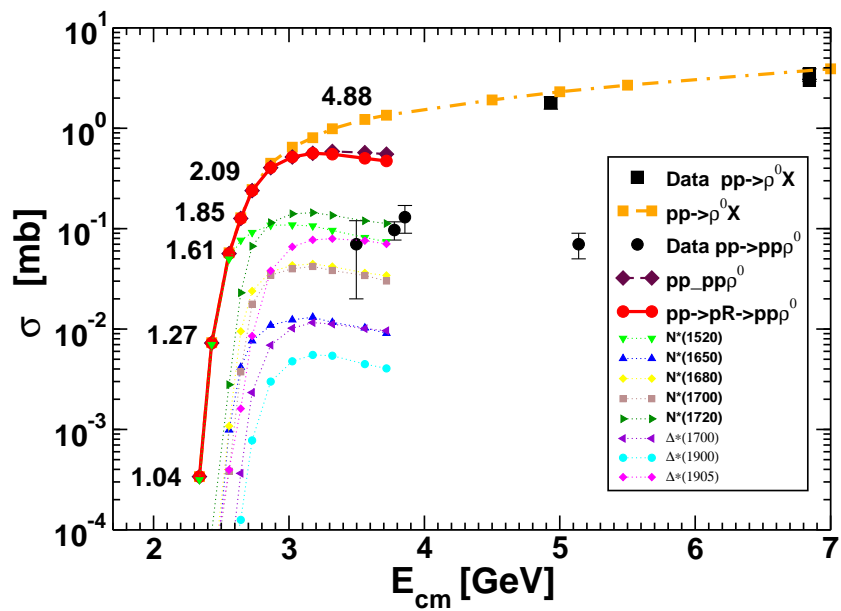}
\caption{(Color online) Cross sections for $\rho^0$ meson production 
in $pp$ collisions. Calculations are shown for inclusive 
($pp\rightarrow \rho^0X$) and 
exclusive ($pp\rightarrow pp\rho^0$) in comparison 
to experimental data \cite{Flaminio:1984gr}. The contribution of the most 
important resonances to the resonant exclusive production is 
additionally shown.} 
\label{rho_inc_v2.3}
\end{figure}

The comparison with DLS data for dilepton production in $pp$ collisions suggested that at the lowest energies 
(1.04-2.09 GeV) the $\rho$ meson yield might be overestimated by our model.
Figure \ref{rho_inc_v2.3} shows the cross sections for the 
inclusive ($pp\rightarrow \rho^0X$) and exclusive ($pp\rightarrow pp\rho^0$) 
production of the neutral $\rho$ meson in $pp$ collisions, 
in comparison with experimental data from Ref. \cite{Flaminio:1984gr}. 
The points corresponding to the energies scanned by the DLS $pp$ program are labelled by the 
corresponding laboratory energies to simplify the readability of the figure. The resonant contribution to the exclusive production, important at the 
energies relevant for this work, is separately shown. 
Moreover, the contribution of the most important resonances is explicitly shown. To specify 
the order of the relative scale, the contribution of some of the less important resonance is shown too. 
The full list of resonances which couple to the $\rho$ meson in the UrQMD model is given in Table \ref{list_res} 
together with the values of the respective  branching ratios in the $N\rho$ decay channel as 
used in UrQMD v2.3. Some of the values for the branching ratios differ from the ones used in UrQMD v1.0 \cite{Bass:1998ca,Ernst:1997yy}. However, the same 
values are used since UrQMD v1.1. Above the threshold 
for meson production by string fragmentation and decay, the $pp\rightarrow pp\rho^0$ 
reaction channel is additionally populated by processes involving strings.

\begin{table}
\caption{List of the non strange resonances included in UrQMD with non vanishing branching ratio 
into the $N\rho$ decay channel.}
	\centering
		\begin{tabular}{|lc||lc|}
\hline\hline
Resonance	\quad	&	  Br($N\rho$) \quad&  Resonance	\quad	&	  Br($N\rho$) \quad \\
\hline
$N^*(1520)$   &  .15    & 
 $\Delta^*(1620)$ & .05   \\ 
$N^*(1650)$   &  .06    &   $\Delta^*(1700)$ & .25   \\
$N^*(1680)$   &  .10    &   $\Delta^*(1900)$ & .25   \\
$N^*(1700)$   &  .20    &   $\Delta^*(1905)$ & .80   \\
$N^*(1710)$   &  .05    &   $\Delta^*(1910)$ & .10   \\
$N^*(1720)$   &  .73    &   $\Delta^*(1930)$ & .22   \\
$N^*(1900)$   &  .15    &   $\Delta^*(1950)$ & .08   \\
$N^*(1990)$   &  .43    &                    &       \\
$N^*(2080)$   &  .12    & 									 &       \\
$N^*(2190)$   &  .24    & 									 &       \\
$N^*(2220)$   &  .22    & 									 &       \\
$N^*(2250)$   &  .25    & 									 &       \\
\hline
		\end{tabular}
\label{list_res}
\end{table}

Unless explicitly specified, in the following we will discuss in terms of laboratory energies.
One observes that in collisions at laboratory energies of 1.04-2.09 GeV the $\rho$ meson 
production is determined by the 
excitation of $\Delta^*$ and $N^*$ resonances in reactions $pp\rightarrow pN^*$ and 
$pp\rightarrow p\Delta^*$ and the inclusive production  of the $\rho$ meson  
coincides with the exclusive production. 
In particular, the latter is practically saturated by the contribution of the $N^*(1520)$ resonance up to beam energies of 1.61 GeV.
On the contrary, at 4.88 GeV, the inclusive 
production dominates by far the exclusive production.
The first datapoints on inclusive production are well reproduced by the model, but 
are far away from the energies spanned by the DLS and the HADES experiments. The exclusive production, 
on the contrary, is systematically overestimated. 

\begin{figure}[hbt]
\includegraphics[height=8cm]{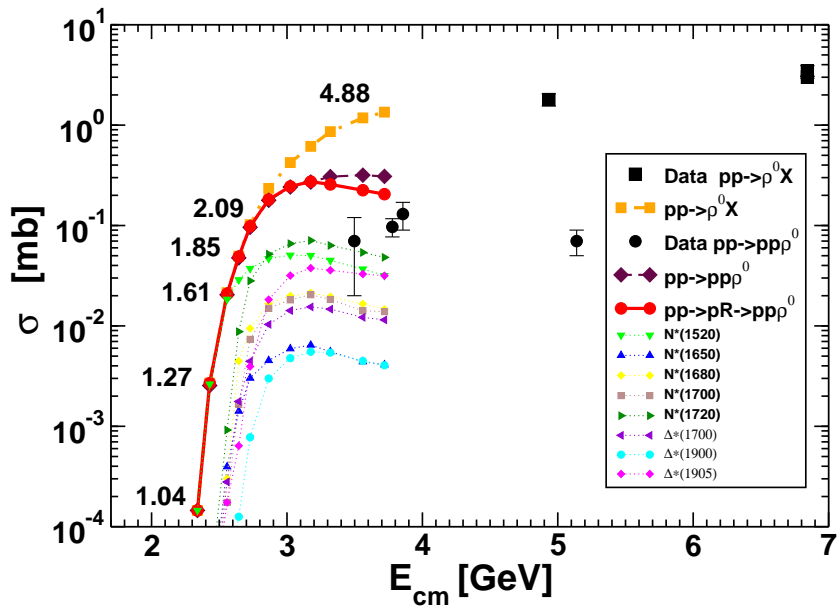}
\caption{(Color online) Same as Fig. \ref{rho_inc_v2.3}, 
but for a smaller value of the $pp\rightarrow p\Delta^*$ and $pp\rightarrow pN^*$ 
cross sections, as explained in the text. }
\label{rho_inc.eps}
\end{figure}

\begin{figure}[h]
\includegraphics[height=16.0cm]{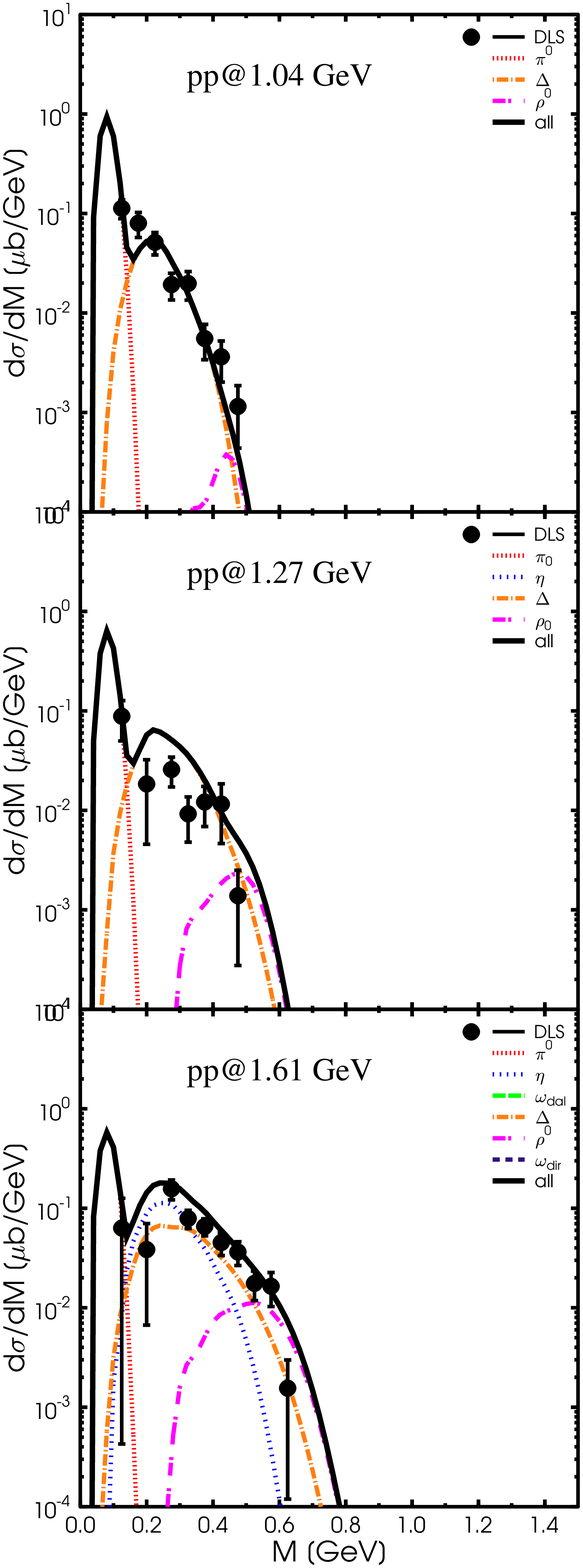}
\hspace*{-.4cm}
\includegraphics[height=16.0cm]{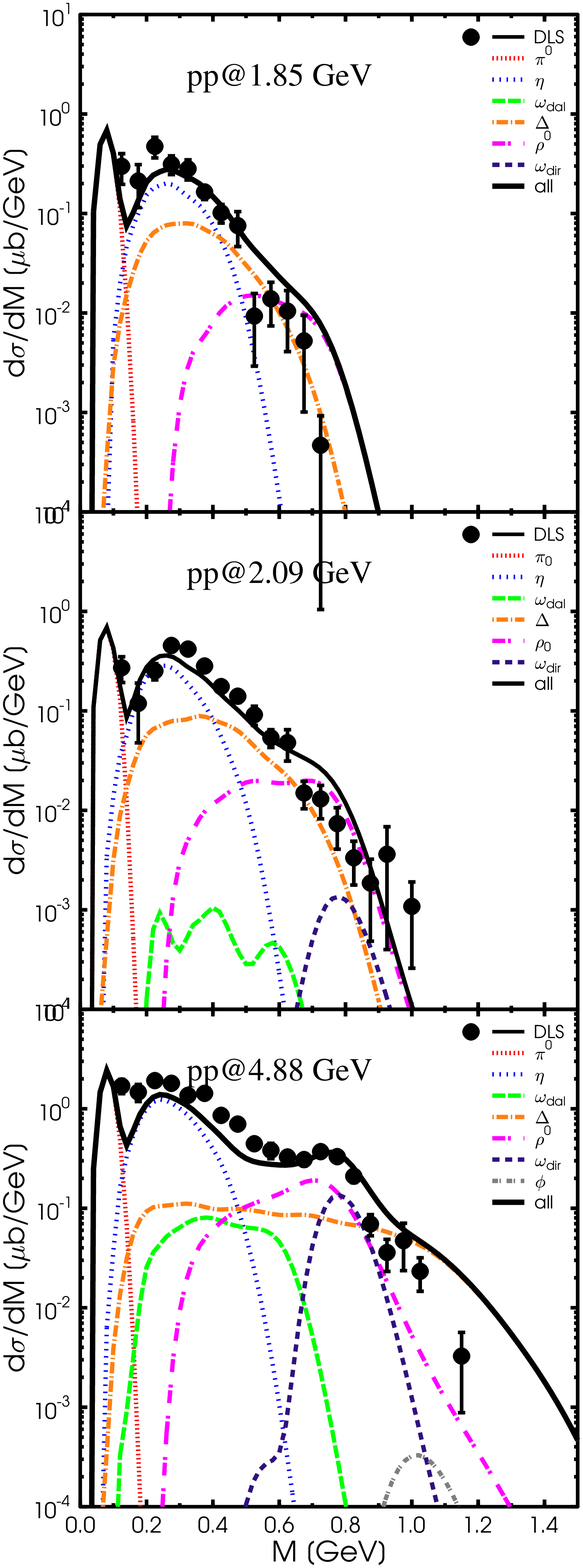}
\caption{(Color online) Same as Fig. \ref{fig:dls_pp}, 
but for a smaller value of the $pp\rightarrow p\Delta^*$ and $pp\rightarrow pN^*$ 
cross sections.}
 \label{fig:dls_pp_nall}
\end{figure}

Poor and often contradictory experimental information is available on the 
production cross sections of $N^*$ and $\Delta^*$ resonances.  For example, in the case of the 
$N^*(1520)$ resonance a reduction of the cross section currently used in UrQMD by a factor 3 is possible in comparison to 
the experimental data \cite{Flaminio:1984gr} and results even in a smaller value of the weighted least mean square. 
For this reason, we investigated the effect that an eventual 
overestimation of the $pp\rightarrow p\Delta^*$ and $pp\rightarrow pN^*$ cross sections would have on 
the $\rho^0$ meson and, consequently, dilepton production. 
Due to the lack of high quality data and to explore the effects of this 
change, we divide all $pp\rightarrow p\Delta^*$ and $pp\rightarrow pN^*$ 
cross sections by a factor 3 with 
exception for the $pp\rightarrow pN^*(1535)$ cross section which is constrained by 
the $\eta$ production. This procedure is surely too crude, but provides a rough estimate of  the consequences that an eventual 
insufficient modelling of the hitherto used 
$pp\rightarrow p\Delta^*$ and $pp\rightarrow pN^*$ 
cross sections might have on 
the model calculations for dilepton spectra.
The results obtained with the modified values of the $pp\rightarrow p\Delta^*$ and $pp\rightarrow pN^*$ 
cross sections are shown in Fig. \ref{rho_inc.eps} and Fig. \ref{fig:dls_pp_nall}. We observe that the model calculations of the exclusive $\rho^0$ meson production 
cross sections moves closer to the experimental data and the DLS data are well described in all mass range. 
In particular, the peak previously observed in the dilepton spectra for $pp$ collisions at 2.09 GeV vanishes to a large extent. 
We notice that the readjustment of the exclusive production of the $\rho^0$ meson 
does not alter the inclusive production at laboratory energy of 4.88 GeV, 
neither the respective result for the dilepton spectra.

\begin{figure}[hbt]
\includegraphics[height=8cm]{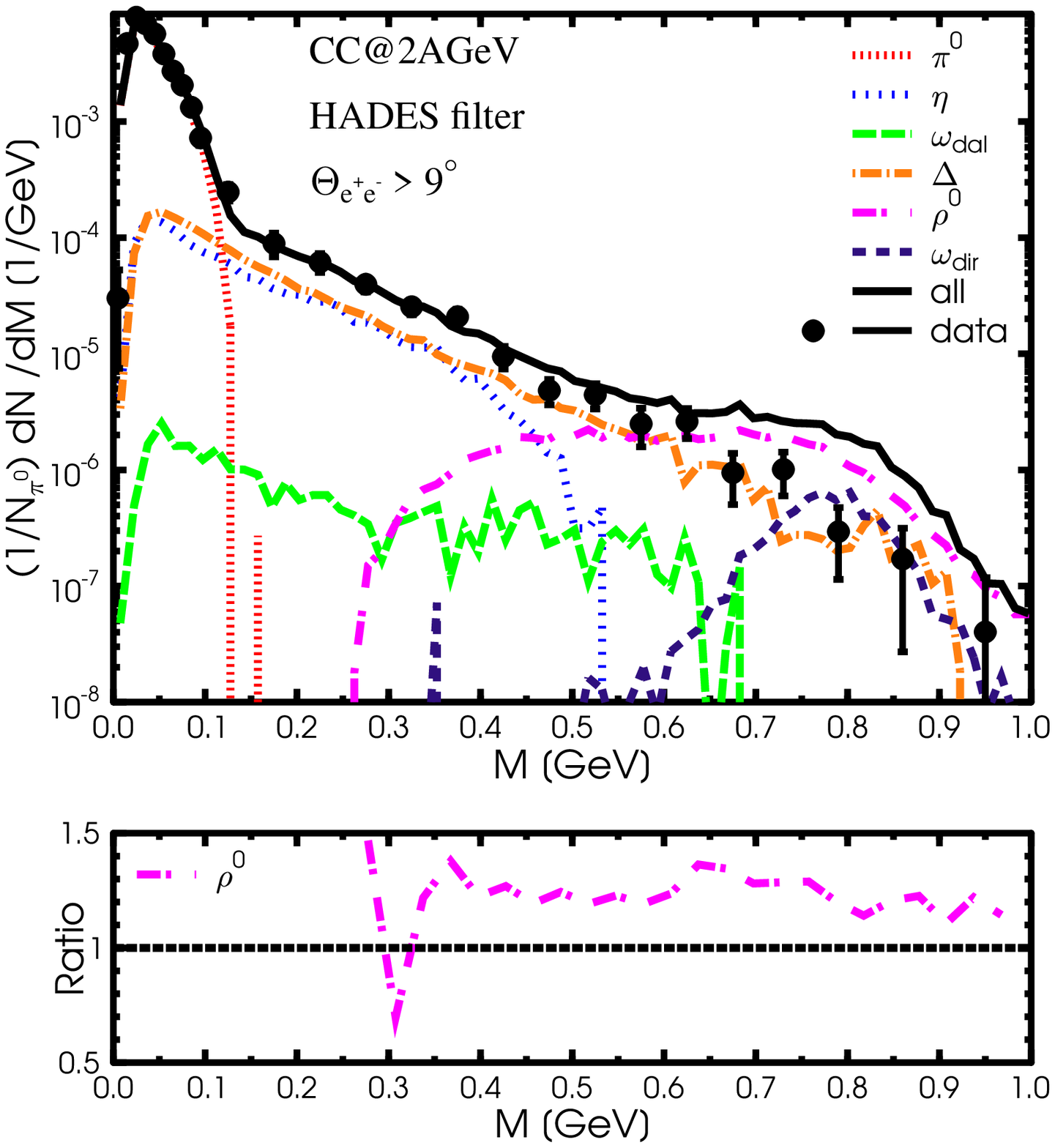}
\caption{(Color online) Upper panel: Same as Fig. \ref{filter_cc2}, 
but for a smaller value of the $pp\rightarrow p\Delta^*$ 
and $pp\rightarrow pN^*$ 
cross sections. Lower panel: Ratio between the $\rho^0$ contribution to the 
dilepton spectra of Fig. \ref{filter_cc2} and Fig. \ref{filter_cc2_shin_nall}. } 
\label{filter_cc2_shin_nall}
\end{figure}

However, the main features of our results remain. In particular, the contribution to the dilepton 
spectrum from $\rho^0$  mesons at the lowest energies, although reduced, is still visible and distinguishable. 
Concerning the reaction C+C at 2 AGeV, we observe that the HADES data remain 
overestimated in the peak region even when the readjusted cross 
sections are used, as shown in Fig. \ref{filter_cc2_shin_nall}.
Many processes, such as multiple scattering, backwards reactions, Fermi motion, etc$\ldots$ distinguish a heavy ion collision from a simple superposition of elementary reactions occurring at the same beam energy. It is also clear that 
in the local equilibrium limit particle production would be statistical 
and information on the employed elementary cross sections would be lost. 
In the present case, which can be seen as an intermediate regime between the two limiting cases of an elementary reaction and an equilibrated system, we find that  a small readjustment of some particular cross sections can still affect the dilepton spectrum, but differences are smaller than in the elementary case.

\section{Predictions for Ar+KCl}
\label{sec:arkcl}
In this section we consider the reaction Ar+KCl at 1.75~AGeV, recently 
measured and currently analyzed by the HADES Collaboration. The predictions 
presented here refer to 
minimum bias calculations  
and have been 
obtained adopting the shining method. 
All spectra are normalised to the pion multiplicity.

\begin{figure}[htb!]
\includegraphics[height=8cm]{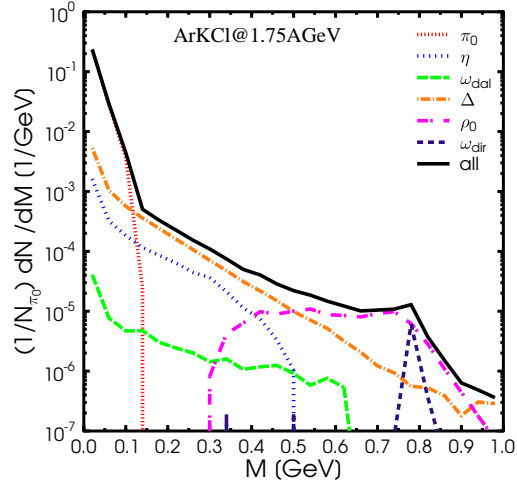}
\caption{(Color online) UrQMD model calculations for dilepton invariant mass spectra from Ar+KCl collisions at beam energy of 1.75~AGeV. The calculations were performed with the shining method.}
\label{fig:arkcl1.75_mass_unfil}
\end{figure}

The invariant mass differential dilepton spectrum is shown in Fig.\ref{fig:arkcl1.75_mass_unfil}.
Compared to C+C at 2~AGeV we observe a smaller contribution of the $\eta$ resonances relatively to the $e^+e^-$-pairs originating from the $\Delta$ Dalitz decay. Up to a dilepton mass of 0.4~GeV the biggest contribution to this mass spectrum occurs from the long-lived mesons $\eta$ and $\pi ^0$ and the baryonic resonance $\Delta$. Considering the contribution originating from vector mesons it is visible that the $\omega$ Dalitz decay again plays only a subordinate role, while the $e^+e^-$-pair production from $\omega$ direct decay becomes important for higher invariant mass, such that in 
the (unfiltered) dilepton spectrum a peak at $M \approx 0.8$~GeV is visible. The direct decay of the vector meson $\rho$ dominates the mass spectrum for $M > 0.5$~GeV.

\section{Tracing  the dilepton emission back in time } \label{sascha}

In this section we investigate the dependence of the dilepton 
signal on the reaction evolution time including the corresponding densities. 
Aim of this analysis is to trace the dilepton emission in time 
to identify the different stages and density regimes of the 
heavy ion collision from which dileptons originate. The study is performed 
for the reaction C+C at 2 AGeV.

\begin{figure}[htb]
\centering
\includegraphics[height=8cm]{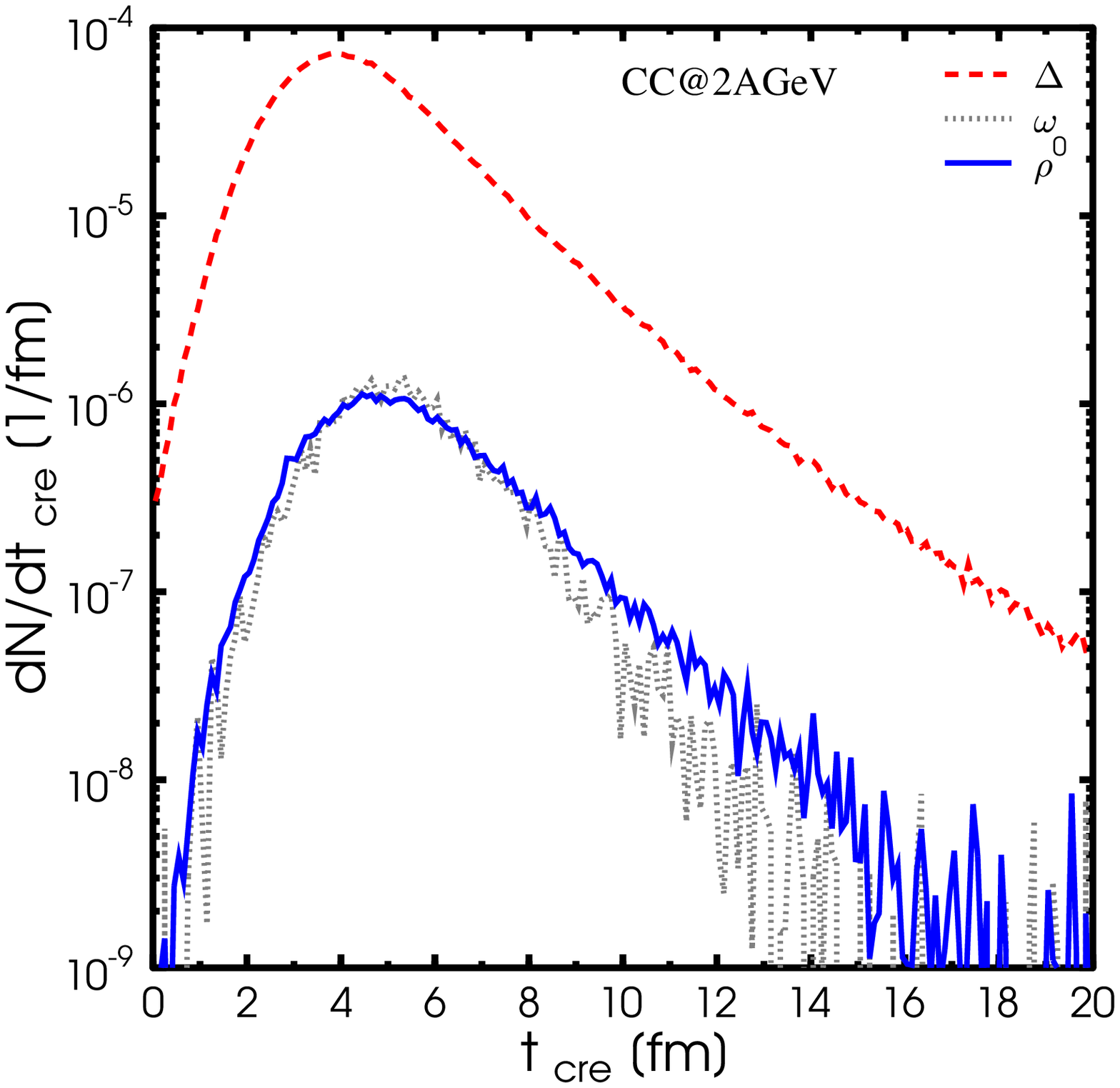}
\hspace*{-1cm}
\includegraphics[height=8cm]{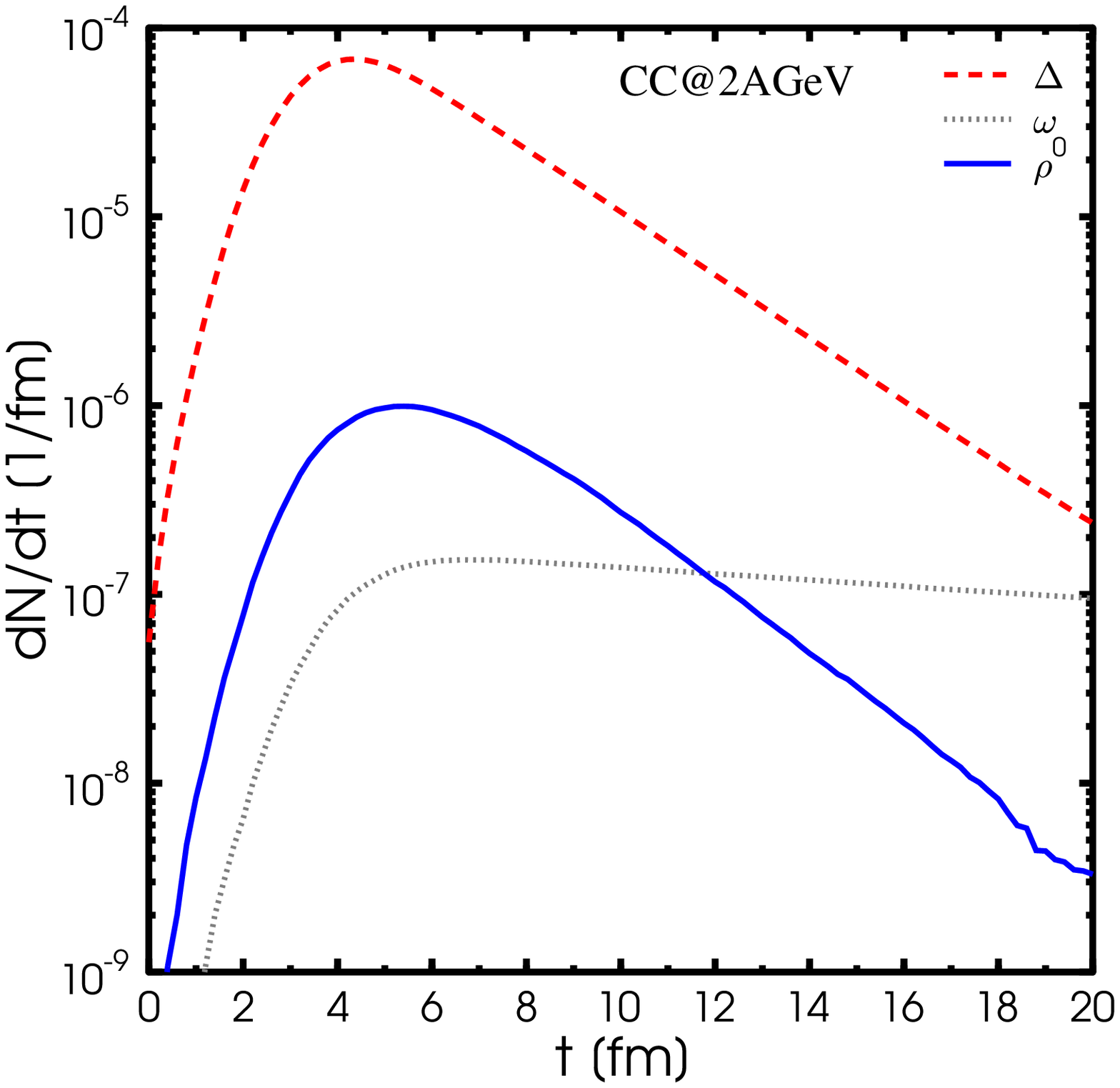}
\caption{(Color online) Dilepton multiplicity 
for minimal bias C+C collisions at beam energies of 2~AGeV 
as a function of the time at which the parent particle made its first 
appearance in the 
evolving system (left panel) and 
corresponding averaged dilepton rate as a function of the evolution time 
of the heavy ion collision (right panel).\label{dn_dt_cre}}
\end{figure}

Let us focus our discussion on the contributions of the vector mesons and the 
$\Delta$ resonance. 
The remaining contributions, $\pi^0$ and $\eta$ Dalitz decays, although large, 
do not play a central role in the physics one aims to explore 
with dilepton experiments and can be viewed as 
some sort of standard ``background''. 
The left panel of Fig.~\ref{dn_dt_cre} shows the dilepton multiplicities as a 
function of the time at 
which the parent particle has been created. In the right panel, 
the multiplicities are shown as a function of the evolution time 
of the heavy ion reaction. In the latter, the continuous emission 
of dileptons from the parent particle 
is explicitly shown, whereas in the former the integrated value is shown.
In other words, from a particle which lives from time $t_i$ 
till time $t_f$, dileptons are emitted with rate
\beq
 \frac{dN^{e^+e^-}(t)}{dt}=
 \begin{cases} 
\Gamma^{e^+e^-}/\gamma &\mbox{for } t_i\leq t \leq t_f \\
 0 &\mbox{otherwise }
\end{cases}
\label{contin_rate}
\eeq
Here $t$ denotes the time in the frame of the evolving system (cm frame of the nucleus-nucleus collision). The Lorentz factor $\gamma$ connects a 
time interval in this system to the corresponding one in the rest frame of 
the emitting particle.
For each particle, the function of $t$ (\ref{contin_rate}) is plotted in the 
right panel of  
Fig.\ref{dn_dt_cre} and 
corresponds to a straight line going from  
$t_i$ to $t_f$. 
The corresponding integral 
\beq
\int_{t_i}^{t_f}\frac{dN^{e^+e^-}(t)}{dt}~dt= \Gamma^{e^+e^-}~\tau
\eeq
where $\tau=(t_f-t_i)/\gamma$ is the life-time of the particle, gives 
the total number of dilepton emitted by the particle (created at $t=t_i$) 
and is reported in the left panel of Fig.\ref{dn_dt_cre}. 

We observe that:
\begin{itemize}
\item Most dileptons originate from particles created within the first 8 fm. 
The emission is maximal from vector mesons created at about 5 fm 
and $\Delta$ resonances created at slightly earlier time (about 3.5 fm). 
This is understandable if one considers that in the resonance model 
vector mesons arise from the decay of baryonic resonances. 
Since the baryonic resonances have a typical total width of the 
order of 100-200 MeV, 
their decay takes typically place about 1-2 fm after their creation. 
\item In the case that the parent particles is a relatively short lived 
particle, e.g. a $\Delta$-resonance or a $\rho$ meson, 
most dileptons are emitted within the first 10 fm, with a maximum 
around 6 fm. Later, for $t>6$ fm, 
the dilepton emission strongly decreases with increasing 
time.  On the contrary, if the parent particles is a long lived particle, 
e.g. a $\omega$ meson, dileptons are emitted continuously at an 
almost constant rate for $t>6$ fm. This is due to the fact that 
those $\omega$ mesons which happened to survive the various absorption 
processes 
live relatively long and emit dileptons during their whole life-time.
\end{itemize}

\begin{figure}[hbt]
\includegraphics[height=16.0cm]{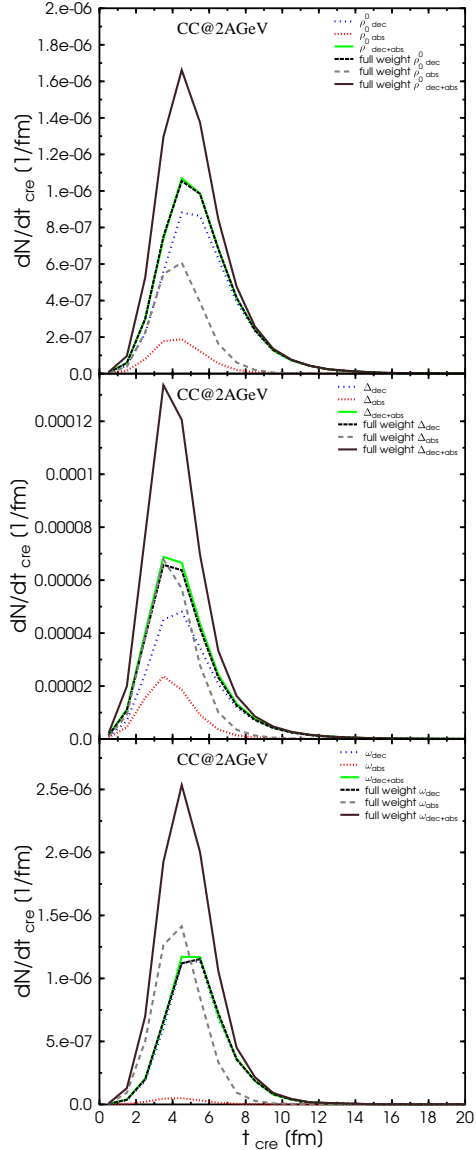}
\caption{(Color online) Dilepton multiplicity 
from minimal bias C+C collisions at beam energies of 2~AGeV 
as a function of the time at which the parent particle made its first 
appearance in the 
evolving system. The dashed lines denote calculation where the full branching ratio into dileptons is attached to both the decay and the absorption vertices. 
\label{shining_time}}
\end{figure}

In Fig.~\ref{shining_time} the role of absorption on the reduction of the 
dilepton signal is shown. The observed yield is compared to the 
yield expected from a vacuum-like picture in which the parent resonance, 
after being produced, does not interact further up to its decay, 
here simply denoted by ``full weight'' scenario. For a detailed 
discussion of the different prescriptions for dilepton production 
see~\cite{Vogel:2007yu}.
The total dilepton signal from vector mesons is reduced by a factor 
1.5(for the $\rho$ meson)-2(for the $\omega$ meson) due to  reabsorption. Especially in the case of the 
$\omega$ meson the ``potential'' dilepton signal of those 
particles which are absorbed (labelled by $\omega_{\rm abs}$ in 
Fig~\ref{shining_time}) is strongly suppressed (by a factor $20$). 

\begin{figure}[hbt]
\includegraphics[height=8cm]{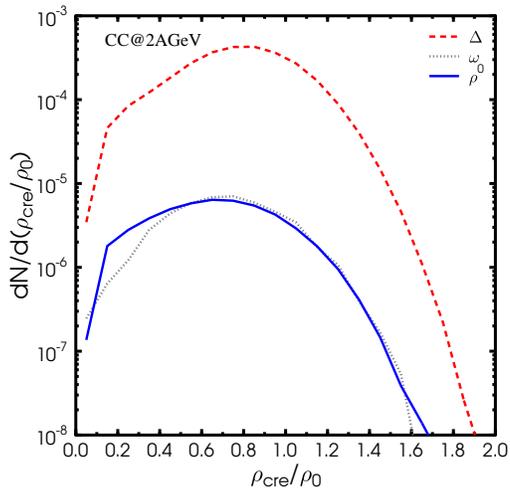}
\caption{(Color online) Dilepton multiplicity 
from minimal bias C+C collisions at beam energies of 2~AGeV 
as a function of the local density present in the space-time point at which 
the parent particle has been created.
\label{dnddens_cre2}}
\end{figure}

\begin{figure}[hbt]
\includegraphics[height=16.0cm]{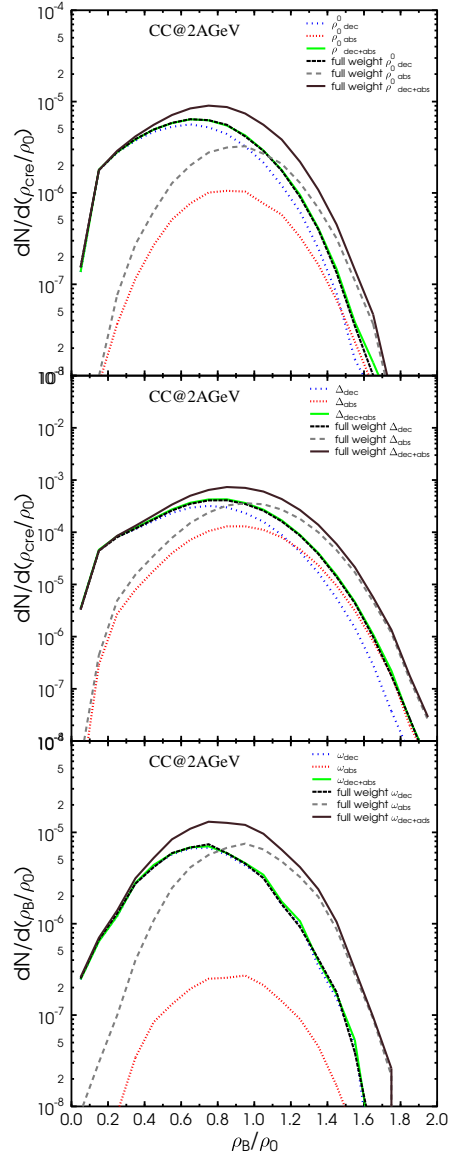}
\caption{(Color online) Dilepton multiplicity 
from minimal bias C+C collisions at beam energies of 2~AGeV 
as a function of the local density present in the space-time point at which the parent particle has been created. The dashed lines denote calculation where the full branching ratio into dileptons is attached to both the decay and the absorption vertices. 
\label{dnddens_cre}}
\end{figure}

Next, we investigate the influence of the 
baryon density locally present on the electromagnetic response of the system, 
as depicted in Fig. \ref{dnddens_cre2}.
It is clear that a particle propagating through a high density zone of the 
system 
will interact, with a certain probability, with the particles 
present in its surroundings. Absorptive interactions, 
e.g. $\rho N \rightarrow N^*(1520)$, will lead 
to the disappearance of the parent particle from the system within shorter 
times 
than its vacuum mean lifetime (determined by its decay width).
As a consequence of its shorter lifetime, the total 
dilepton yield from the particle will be reduced with respect to the 
yield expected if the particle would be present in the system until its 
decay and emit dileptons for a time interval $\tau_{dec}$.
In particular, the number of dileptons expected to be emitted by a parent 
particle created in a space-time point characterised by a local 
baryon density $\rho_{cre}$ is analysed. The result is 
reported in Fig.\ref{dnddens_cre}.

We observe that between 13\% and 20\% of dileptons originate from particles 
created at densities $\rho_{cre}>\rho_0$ and that absorption reduces 
the potential dilepton yield from these particles by a factor 1.5. 
This effect is particularly strong in the case of the $\omega$ meson.
It is evident from the previous analysis that the parent particles seem 
to be characterised by relatively short lifetimes in the high density phase.

\section{Summary and conclusion} \label{concl}
Dilepton production in nucleus-nucleus and proton-proton reactions at 
SIS/BEVELAC energies has been analysed within the microscopic  transport model UrQMD. 
The results for invariant mass differential dilepton spectra have been compared to 
HADES data for C+C collisions at 1 AGeV and 2 AGeV and to DLS data for $pp$ reactions.
Additionally, predictions for dilepton spectra in $pp$ reactions at 1.25 GeV, 2.2 GeV and 3.5 GeV 
as well as in Ar+KCl at 1.78 AGeV have been presented.
The analysis shows that the low mass region of the dilepton spectra for C+C collisions 
is slightly underestimated by the model calculations at 1 AGeV, but well described at 2AGeV.

The dilepton emission was analyzed in dependence of the evolution time and densities 
typical for the regime probed by the HADES experiment. In particular, the influence of absorption of the parent 
resonances on their dilepton emission has been discussed. We found that absorption is responsible for a global suppression of the dilepton signal of about a 
factor 
1.5-2. 
The absorption processes are more copious in the 
high density phase, resulting in a stronger suppression for particles (and therefore dileptons) produced at the 
highest densities.

\section{Acknowledgements}
We thank Y. Pachmayer 
for providing the experimental data and the HADES filter for the reaction C+C at 1 AGeV, and E.~L.~Bratkovskaya for discussions on the $\rho$ meson production cross section.
This work was supported by BMBF, GSI and the Hessen Initiative 
for Excellence (LOEWE) through the Helmholtz International Center for FAIR (HIC for FAIR).


\begin{thebibliography}{99}


\bibitem{Mazzoni:1994rb}
  M.~A.~Mazzoni  [HELIOS/3 Collaboration.],
  Nucl.\ Phys.\  A {\bf 566}, 95C (1994);
  M.~Masera  [HELIOS Collaboration],
  Nucl.\ Phys.\  A {\bf 590}, 93C (1995).
  

\bibitem{Li:1994cj}
  G.~Q.~Li and C.~M.~Ko,
  Nucl.\ Phys.\  A {\bf 582}, 731 (1995).
  
  
\bibitem{Agakishiev:1995xb}
  G.~Agakishiev {\it et al.}  [CERES Collaboration],
  Phys.\ Rev.\ Lett.\  {\bf 75}, 1272 (1995);
  A.~Drees,
  Nucl.\ Phys.\  A {\bf 610}, 536C (1996).


\bibitem{Li:1995qm}
  G.~Q.~Li, C.~M.~Ko and G.~E.~Brown,
  Phys.\ Rev.\ Lett.\  {\bf 75}, 4007 (1995).

\bibitem{Rapp:1995zy}
  R.~Rapp, G.~Chanfray and J.~Wambach,
  Phys.\ Rev.\ Lett.\  {\bf 76}, 368 (1996);
  R.~Rapp and J.~Wambach,
  Adv.\ Nucl.\ Phys.\  {\bf 25}, 1 (2000).


\bibitem{Ko:1996is}
  C.~M.~Ko, G.~Q.~Li, G.~E.~Brown and H.~Sorge,
  Nucl.\ Phys.\ A {\bf 610}, 342C (1996).

\bibitem{Li:1996mi}
  G.~Q.~Li, C.~M.~Ko, G.~E.~Brown and H.~Sorge,
  Nucl.\ Phys.\  A {\bf 611}, 539 (1996).

\bibitem{Rapp:1997fs}
  R.~Rapp, G.~Chanfray and J.~Wambach,
  Nucl.\ Phys.\  A {\bf 617}, 472 (1997).

\bibitem{Friman:1997tc}
  B.~Friman and H.~J.~Pirner,
  Nucl.\ Phys.\  A {\bf 617}, 496 (1997).


\bibitem{Cassing:1997jz}
  W.~Cassing, E.~L.~Bratkovskaya, R.~Rapp and J.~Wambach,
  Phys.\ Rev.\  C {\bf 57}, 916 (1998).


\bibitem{Porter:1997rc}
  R.~J.~Porter {\it et al.}  [DLS Collaboration],
  Phys.\ Rev.\ Lett.\  {\bf 79}, 1229 (1997).


\bibitem{Bratkovskaya:1997mp}
  E.~L.~Bratkovskaya, W.~Cassing, R.~Rapp and J.~Wambach,
  Nucl.\ Phys.\  A {\bf 634}, 168 (1998).

\bibitem{Ernst:1997yy}
 C.~Ernst, S.~A.~
 s, M.~Belkacem, H.~Stoecker and W.~Greiner,
  Phys.\ Rev.\ C {\bf 58}, 447 (1998).

\bibitem{Shekhter:2003xd}
  K.~Shekhter, C.~Fuchs, A.~Faessler, M.~Krivoruchenko and B.~Martemyanov,
  Phys.\ Rev.\  C {\bf 68}, 014904 (2003).


\bibitem{Arnaldi:2006jq}
  R.~Arnaldi {\it et al.}  [NA60 Collaboration],
  Phys.\ Rev.\ Lett.\  {\bf 96}, 162302 (2006).


\bibitem{Adamova:2006nu}
  D.~Adamova {\it et al.},
  Phys.\ Lett.\  B {\bf 666}, 425 (2008).




\bibitem{Agakichiev:2006tg}
  G.~Agakichiev {\it et al.}  [HADES Collaboration],
  Phys.\ Rev.\ Lett.\  {\bf 98}, 052302 (2007).




\bibitem{Agakishiev:2007ts}
  G.~Agakishiev {\it et al.}  [HADES Collaboration],
  Phys.\ Lett.\  B {\bf 663}, 43 (2008).




\bibitem{Cozma:2006vp}
  M.~D.~Cozma, C.~Fuchs, E.~Santini and A.~Fassler,
  Phys.\ Lett.\  B {\bf 640}, 170 (2006).
  
\bibitem{Schumacher:2006wc}
  D.~Schumacher, S.~Vogel and M.~Bleicher,
  Acta Phys.\ Hung.\  A {\bf 27}, 451 (2006).
 
\bibitem{Thomere:2007cj}
  M.~Thomere, C.~Hartnack, G.~Wolf and J.~Aichelin,
  Phys.\ Rev.\  C {\bf 75}, 064902 (2007).


\bibitem{Ruppert:2007cr}
  J.~Ruppert, C.~Gale, T.~Renk, P.~Lichard and J.~I.~Kapusta,
  Phys.\ Rev.\ Lett.\  {\bf 100}, 162301 (2008).


\bibitem{Bratkovskaya:2007jk}
  E.~L.~Bratkovskaya and W.~Cassing,
  Nucl.\ Phys.\  A {\bf 807}, 214 (2008).


\bibitem{Vogel:2007yu}
  S.~Vogel, H.~Petersen, K. Schmidt, E. Santini, C. Sturm, J.~Aichelin and M.~Bleicher,
  Phys.\ Rev.\  C {\bf 78}, 044909 (2008).
  
\bibitem{Santini:2008pk}
  E.~Santini, M.~D.~Cozma, A.~Faessler, C.~Fuchs, M.~I.~Krivoruchenko and B.~Martemyanov,
  Phys.\ Rev.\  C {\bf 78}, 034910 (2008).



\bibitem{Brown:1991kk}
  G.~E.~Brown and M.~Rho,
  Phys.\ Rev.\ Lett.\  {\bf 66}, 2720 (1991).


\bibitem{Hatsuda:1991ez}
  T.~Hatsuda and S.~H.~Lee,
  Phys.\ Rev.\  C {\bf 46}, 34 (1992).


\bibitem{Peters:1997va}
  W.~Peters, M.~Post, H.~Lenske, S.~Leupold and U.~Mosel,
  Nucl.\ Phys.\  A {\bf 632}, 109 (1998);
  M.~Post, S.~Leupold and U.~Mosel,
  Nucl.\ Phys.\  A {\bf 689}, 753 (2001);
  M.~Post, S.~Leupold and U.~Mosel,
  Nucl.\ Phys.\  A {\bf 741}, 81 (2004).


\bibitem{Lutz:2001mi}
  M.~F.~M.~Lutz, G.~Wolf and B.~Friman,
  Nucl.\ Phys.\ A {\bf 706}, 431 (2002)
  [Erratum-ibid.\ A {\bf 765}, 431 (2006)].


  
\bibitem{aichelin86a}
J.~Aichelin and H.~St\"ocker,
\newblock Phys. Lett. {\bf B176}, 14 (1986);
H.~Sorge, H.~St\"ocker, and W.~Greiner,
\newblock Annals of Physics {\bf 192}, 266 (1989);
J.~Aichelin,
\newblock Phys. Rep. {\bf 202}, 233 (1991).

\bibitem{bass95c}
S.~A. Bass, C.~Hartnack, H.~St\"ocker, and W.~Greiner,
\newblock Phys. Rev. {\bf C51}, 3343 (1995).


\bibitem{winckelmann96a}
L.~A. Winckelmann {\em et~al.},
Nucl.\ Phys.\ A {\bf 610}, 116c (1996).



\bibitem{Sturm:2000dm}
  C.~Sturm {\it et al.}  [KaoS Collaboration],
  Phys.\ Rev.\ Lett.\  {\bf 86}  39 (2001).
  

\bibitem{Bass:1998ca}
  S.~A.~Bass {\it et al.},
  Prog.\ Part.\ Nucl.\ Phys.\  {\bf 41}, 225 (1998).

\bibitem{Bleicher:1999xi}
  M.~Bleicher {\it et al.},
  J.\ Phys.\ G {\bf 25} 1859 (1999).

\bibitem{Petersen:2008kb}
  H.~Petersen, M.~Bleicher, S.~A.~Bass and H.~Stocker,
  arXiv:0805.0567 [hep-ph].


\bibitem{winckelmann95a}
L.~A. Winckelmann, H.~Sorge, H.~St\"ocker, and W.~Greiner,
\newblock Phys. Rev. {\bf C51}, R9 (1995).


\bibitem{PDG06}
  W.~M.~Yao {\it et al.}  [Particle Data Group],
  J.\ Phys.\ G {\bf 33}, 1 (2006).


\bibitem{Holzmann:1997mu}
  R.~Holzmann {\it et al.}  [TAPS Collaboration],
  Phys.\ Rev.\  C {\bf 56}, 2920 (1997).

\bibitem{Calen:1998vh}
  H.~Calen {\it et al.},
  Phys.\ Rev.\  C {\bf 58}, 2667 (1998).
  
 

\bibitem{Teis:1996kx}
  S.~Teis, W.~Cassing, M.~Effenberger, A.~Hombach, U.~Mosel and G.~Wolf,
  Z.\ Phys.\  A {\bf 356}, 421 (1997).
  
\bibitem{Averbeck:1997ma}
  R.~Averbeck {\it et al.}  [TAPS Collaboration],
  Z.\ Phys.\  A {\bf 359} (1997) 65.

\bibitem{Landsberg:1986fd}
  L.~G.~Landsberg,
  Phys.\ Rept.\  {\bf 128}, 301 (1985).

\bibitem{Koch:1992sk}
  P.~Koch,
  Z.\ Phys.\ C {\bf 57}, 283 (1993).

\bibitem{Faessler:1999de}
  A.~Faessler, C.~Fuchs and M.~I.~Krivoruchenko,
  Phys.\ Rev.\  C {\bf 61}, 035206 (2000).

\bibitem{Wolf:1990ur}
  G.~Wolf, G.~Batko, W.~Cassing, U.~Mosel, K.~Niita and M.~Schaefer,
  Nucl.\ Phys.\ A {\bf 517}, 615 (1990)

\bibitem{Bratkovskaya:1999mr}
  E.~L.~Bratkovskaya, W.~Cassing, M.~Effenberger and U.~Mosel,
  Nucl.\ Phys.\  A {\bf 653}, 301 (1999).

\bibitem{Heinz:1991fn}
  U.~W.~Heinz and K.~S.~Lee,
  Nucl.\ Phys.\  A {\bf 544}, 503 (1992).

 
\bibitem{Wilson:1997sr}
  W.~K.~Wilson {\it et al.}  [DLS Collaboration],
  Phys.\ Rev.\  C {\bf 57}, 1865 (1998).

\bibitem{Faessler:2000md}
  A.~Faessler, C.~Fuchs, M.~I.~Krivoruchenko and B.~V.~Martemyanov,
  J.\ Phys.\ G {\bf 29}, 603 (2003).

\bibitem{Bratkovskaya:1998pr}
  E.~L.~Bratkovskaya and C.~M.~Ko,
  Phys.\ Lett.\  B {\bf 445}, 265 (1999).


\bibitem{Kaptari:2005qz}
  L.~P.~Kaptari and B.~Kampfer,
  Nucl.\ Phys.\  A {\bf 764}, 338 (2006).
  
  
\bibitem{Schafer:1989dm}
  M.~Schafer, T.~S.~Biro, W.~Cassing and U.~Mosel,
  Phys.\ Lett.\  B {\bf 221}, 1 (1989).
  
\bibitem{Shyam:2003cn}
  R.~Shyam and U.~Mosel,
  Phys.\ Rev.\  C {\bf 67}, 065202 (2003).
 
\bibitem{Shyam:2008rx}
  R.~Shyam and U.~Mosel,
  arXiv:0811.0739 [hep-ph].

\bibitem{Flaminio:1984gr}
  V.~Flaminio, W.~G.~Moorhead, D.~R.~O.~Morrison and N.~Rivoire, 
  Compilation of cross. sections III: $p$ and $\bar{p}$ induced reactions, 
  CERN-HERA 84-01 (1984). 

\end{thebibliography}
\end{document}